%% file: main.tex
\documentclass[superscriptaddress,
aps,
prl,
twocolumn,
showpacs,
nofootinbib,
longbibliography,
notitlepage,
floatfix]{revtex4-2}

\usepackage{mathtools}
\usepackage{amsbsy}
\usepackage{amstext}
\usepackage{amsmath}
\usepackage{amssymb}
\usepackage{txfonts}
\usepackage{color}
\usepackage{graphicx}
\usepackage{wrapfig}
\usepackage{float}
\usepackage[normalem]{ulem}
\usepackage{physics}
\usepackage{bbold}
\usepackage{cancel}

\usepackage[version=4]{mhchem} 
\usepackage{xcolor} 
\usepackage{multirow} 
\usepackage[caption=false]{subfig}

\usepackage[unicode=true,pdfusetitle,
 bookmarks=false,
 breaklinks=false,
 pdfborder={0 0 1},
 backref=false,
 colorlinks=true, linkcolor=magenta, urlcolor=gray, citecolor=olive, filecolor=blue]
 {hyperref}

\hypersetup{
 bookmarksnumbered=false,
 bookmarksopen=false}

\makeatletter

\@ifundefined{textcolor}{}{
 \definecolor{BLACK}{gray}{0}
 \definecolor{WHITE}{gray}{1}
 \definecolor{RED}{rgb}{1,0,0}
 \definecolor{GREEN}{rgb}{0,0.7,0}
 \definecolor{BLUE}{rgb}{0,0,1}
 \definecolor{CYAN}{cmyk}{1,0,0,0}
 \definecolor{MAGENTA}{cmyk}{0,1,0,0}
 \definecolor{YELLOW}{cmyk}{0,0,1,0}
}

\newcommand{\beq}{\begin{equation}}
\newcommand{\eeq}{\end{equation}}
\newcommand{\beqa}{\begin{eqnarray}}
\newcommand{\eeqa}{\end{eqnarray}}

\let\origaddcontentsline\addcontentsline
\newcommand{\suppressTOC}{\let\addcontentsline\@gobblethree}
\newcommand{\restoreTOC}{\let\addcontentsline\origaddcontentsline}

\makeatother
\begin{document}

\preprint{APS/123-QED}

\suppressTOC

\title{Suppressing Fast Dipolar Noise in Solid-State Spin Qubits}

\author{Jaime García Oliván}
\affiliation{Department of Physical Chemistry, University of the Basque Country UPV/EHU, Apartado 644, 48080 Bilbao, Spain}
\affiliation{EHU Quantum Center, University of the Basque Country UPV/EHU, Bilbao, Spain}

\author{Ainitze Biteri-Uribarren}
\affiliation{Department of Physical Chemistry, University of the Basque Country UPV/EHU, Apartado 644, 48080 Bilbao, Spain}
\affiliation{EHU Quantum Center, University of the Basque Country UPV/EHU, Bilbao, Spain}

\author{Oliver T. Whaites}
\affiliation{Department of Physical Chemistry, University of the Basque Country UPV/EHU, Apartado 644, 48080 Bilbao, Spain}
\affiliation{EHU Quantum Center, University of the Basque Country UPV/EHU, Bilbao, Spain}

\author{Jorge Casanova}
\affiliation{Department of Physical Chemistry, University of the Basque Country UPV/EHU, Apartado 644, 48080 Bilbao, Spain}
\affiliation{EHU Quantum Center, University of the Basque Country UPV/EHU, Bilbao, Spain}

\date{\today}

\begin{abstract}

Spin qubit coherence is a fundamental resource for the realization of quantum technologies. For solid-state platforms,  spin decoherence is dominated by the magneto-active environment in the lattice, limiting their applicability. While standard dynamical decoupling techniques, such as the Hahn echo, extend central spin coherence, they fail to suppress the fast noise arising from strong dipolar interactions within the bath. Here, we present a decoupling mechanism, Hybrid-LG, that suppresses intra-bath dipolar interactions --thus, fast noise acting on spin qubits-- and demonstrate its effectiveness in extending spin coherence through efficient in-house CCE simulations. Specifically, we investigate one of the most widely exploited solid-state quantum platforms: an ensemble of nitrogen-vacancy (NV) centers in diamond coupled to a large and dense bath of substitutional nitrogen paramagnetic impurities (P1 centers). Our results reveal at least a twofold enhancement in NV coherence time relative to standard techniques including P1 center driving, without requiring additional control power.

\end{abstract}

\maketitle

Solid-state spin defects have shown promising developments as quantum bits in numerous applications of quantum technologies \cite{Wolfowicz_2021,Liu_2022}, ranging from high-precision quantum sensing \cite{DegenCapellaroReview,CastellettoColorCenters,cAERIS,Jens_2024,Grafenstein_2025,Rizzato_2025} to quantum computing and simulations \cite{Zwerver_2023,Carlos_Q_Sim,Randall_Simulator,Blind_QC,Abobeih_2022}. These qubits typically benefit from optical accessibility and room temperature operability, and provide a route towards scalable quantum technologies. In this scenario, extending qubit coherence time would enable distinct key applications, such as high-fidelity quantum gates~\cite{Abobeih_2022,Huang_2024,Simmons_2024}, enhanced sensitivity magnetometers~\cite{Hong_2013,Barry2020sensitivity,Du_2024} and secure quantum communications ~\cite{Herbschleb_2019,Hermans_2022}, among others.

While the host lattice shields spin qubits from external noise, intrinsic noise sources within the material still limit their coherence. In particular, the principal dephasing mechanism is often attributed to the magneto-active bath of spins surrounding a qubit \cite{DegenCapellaroReview,Bauch2020,onizhukCompPersp,BelthangadyNVP1}, whose noise dynamics are dictated by the bath's properties. For sparse environments, weak intra-bath dipolar couplings lead to slowly fluctuating magnetic noise fields, whereas in dense environments, strong intra-bath couplings generate fast noise fluctuations, particularly when the bath spins are electron-like.

Many methods to minimize spin bath noise have been developed, such as isotopic purification for nuclear noise suppression and dynamical decoupling (DD) protocols \cite{Carlos_DD,Han_DD,Rizzato_DD,Ezzell_DD}. The latter have proven to be particularly successful in achieving longer qubit coherence times by refocusing slow or static noise, exceeding $T_2 \sim 100 \mathrm{\ \mu s}$ with the simple Hahn echo sequence \cite{Bauch2020,Barry2024}. Nonetheless, these protocols fail to decouple the fast electronic noise arising from strongly coupled spin baths \cite{Chrostoski_2021,Bauch2020,ZhaoNV,WitzelSpaguetti}, presenting a bottleneck for the realization of these quantum technological platforms.

Significant effort has been channeled into suppressing this residual electronic noise by modifying material properties \cite{Zheng_2022,Hao_2025,KiLo_2025,SchatzlePCCE,MarcksCCE}, e.g. through sample surface treatments and termination. However, such approaches do not address the dominant sources of electronic noise in the bulk. Other experimental studies have instead focused on actively driving the spin bath, either stroboscopically or continuously, to enhance the coherence time of the qubit \cite{knowles-bathDriving,Bauch2018,Joos_2022,BelthangadyNVP1}. While this strategy has shown promise, achieving sufficient driving power to effectively decouple the bath as well as addressing electronic systems with multiple resonance lines remains technically challenging and introduces adverse heating. To the best of our knowledge, this process has not previously been studied or optimized, likely due to the significant challenges apparent in modeling large systems.

\begin{figure}[b]
    \centering
    \includegraphics[width=0.4\textwidth]{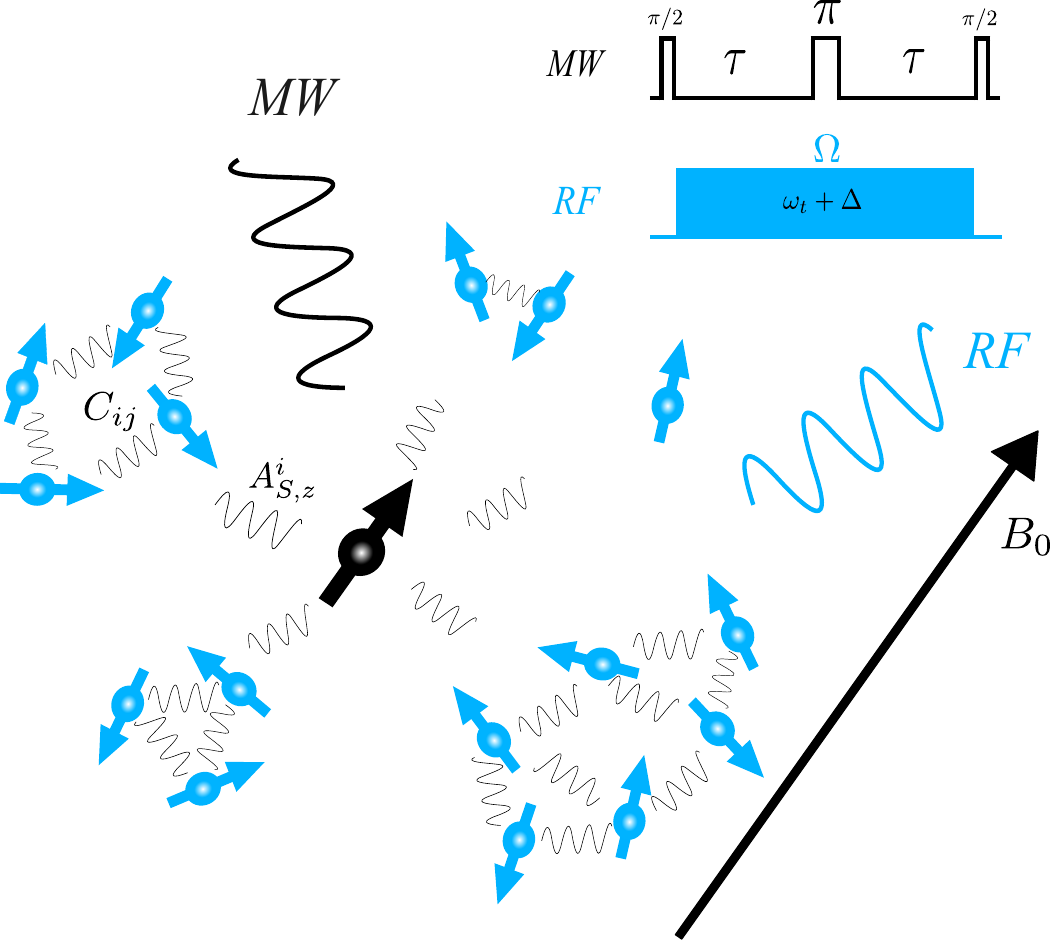}
    \caption{Schematic of the system. A central spin (in black) is surrounded by a magnetic environment composed of \textit{clusters} of strongly interacting spins (in blue). A Hahn echo sequence is applied to the central spin while a driving field of strength $\Omega$, which can be detuned by an amount $\Delta$ from the bath resonant (target) frequency $\omega_t$, decouples the environment.}
    \label{F: System}
\end{figure}

In this Letter, we develop a general and efficient numerical framework to model driven bath dynamics, grounded in the gold standard for quantum many-body simulations --the Cluster Correlation Expansion (CCE) \cite{CCE-1,CCE-2}. A notable advance of our numerical procedure is the incorporation of a mean-field theory for the driven spin bath, which captures collective bath effects --essential for accurate simulations of bath decoupling protocols. Leveraging our model, we investigate a key solid-state quantum platform: an ensemble of NV center qubits coupled to a dense bath of P1 centers \cite{DegenDarkSpins,Bartling2025, WitzelSpaguetti, SchatzlePCCE, BelthangadyNVP1}, often identified as the main source of NV dephasing \cite{Bauch2020}. Building on these insights, we introduce \textit{Hybrid-LG}, a decoupling protocol for strongly coupled baths of paramagnetic impurities with complex energy spectra in solid-state systems, based on the simultaneous and continuous application of resonant and detuned driving fields. Our method suppresses intra-bath interactions and narrows noise dispersion, facilitating coherence refocusing through standard DD over the qubit. We find our decoupling scheme to yield an enhancement in the NV coherence time by at least a factor of two with respect to standard resonant driving; crucially, with no additional power input \cite{knowles-bathDriving,Bauch2018,Joos_2022,BelthangadyNVP1}.

Our underlying framework consists of an ensemble of qubits, or central spins, in a solid-state lattice hosting a strongly coupled spin bath (see Fig. \ref{F: System}) with concentration $\rho$ -in parts-per-million- (ppm). We assume that the lattice is sparsely populated by central spins, so that those in the ensemble are independent. A strong bias magnetic field is applied along the $z$-axis, leading to a pure dephasing interaction between the central spin and the bath:
\begin{equation}\label{Eq: SecInt}
    \mathcal{H}_{SB} = S_z \sum_{i=1}^N A_{S, z}^i J_z^i,
\end{equation}
where $N$ is the total number of spins in the bath, $S_z$ ($J_z^i$) is the projection of the central spin ($i$-th bath spin) operator along the direction of the field in a two-level basis $\{|u\rangle, |d\rangle\}$ and $A_{S, z}^i$ is the secular coupling between them. The interaction in Eq. (\ref{Eq: SecInt}) effectively describes the coupling of the central spin to a noisy magnetic field. As is common, we apply microwave (MW) pulses to the central spin in order to filter out quasi-static noise, where these pulses are described by the Hamiltonian $\mathcal{H}_p = \Omega_{MW}(t) S_\phi$, defining $\Omega_{MW}(t)$ and $\phi$ as the waveform and the pulse axis, respectively. In particular, we study the evolution of the central spin under the well-known Hahn echo sequence, detailed in Fig. \ref{F: System}. However, the dipolar interaction between bath spins generate random, fast time-dependent dynamics in Eq. (\ref{Eq: SecInt}) leading to dephasing of the central spin, which standard echo-based techniques cannot suppress. Specifically, the intra-bath dipolar coupling is given by
\begin{equation}\label{Eq: DipInt}
    \mathcal{H}_{BB} = \sum_{i=1}^N\sum_{j<i}^N C_{ij} \left[J_z^i J_z^j - \frac{1}{2}(J_x^i J_x^j + J_y^i J_y^j)\right],
\end{equation}
where $\mathbf{J}^i$ is the $i$-th bath spin operator and $C_{ij} \propto \gamma_i \gamma_j / r_{ij}^3$ is the interaction coupling between the $i$-th and $j$-th bath spins, with $\gamma_i$ and $\gamma_j$ their gyromagnetic ratios and $r_{ij}$ their relative distance. The off-axis terms in Eq. (\ref{Eq: DipInt}) give rise to a rapid flip-flopping of the bath spins --the origin of the time dependence of the noise field. For a more thorough description of spin bath induced decoherence, see Supplementary Material, Section \ref{SM: Decoherence}.

In order to improve the Hahn echo coherence time of the central spin, we may add a spin bath driving of the form (in the rotating frame of the bath, i.e., with respect to $\mathcal{H}_0 = \sum_{i=1}^N \omega_t J_z^i$):
\begin{equation}\label{Eq: BathDriving}
    \mathcal{H}_D = \sum_{i=1}^N \Omega J_\alpha^i + \Delta J_z^i,
\end{equation}
where $\Delta = \omega_D - \omega_t$ is a frequency detuning of the driving, $\omega_D$ ($\omega_t$) being the driving (target) frequency and $J_\alpha^i = \sin\alpha J_x^i + \cos\alpha J_y^i$, with $\alpha$ its initial phase. When the bath driving is resonant (that is, when $\Delta = 0$) the bath spins undergo Rabi oscillations, suppressing the interaction in Eq. \eqref{Eq: SecInt} as long as $\Omega \gg |A_{S, z}|$. This effect has been previously studied and demonstrated experimentally \cite{knowles-bathDriving,Bauch2018,Joos_2022,BelthangadyNVP1}. When the driving is detuned, the Hamiltonian~(\ref{Eq: BathDriving}) can be rewritten as $\mathcal{H}_D = \sum_{i=1}^N \bar{\Omega} J_P^i$, where $\bar{\Omega} \equiv \sqrt{\Omega^2 + \Delta^2}$ is the effective Rabi frequency of the driving and $P$ is the effective rotation axis. Interestingly, if the frequency detuning satisfies the so-called LG \textit{magic} condition \cite{Lee-Goldburg}, that is where $\Delta = \pm \Omega / \sqrt{2}$, the dipolar interactions in Eq. \eqref{Eq: DipInt} are canceled (up to first order) if $\bar{\Omega} \gg |C_{ij}|$, reducing the frequency dispersion of the noise (see Supplementary Information, Section \ref{SM: Protocols}).

Spin baths in solid-state systems are often composed of crystal defects that comprise of both an electron and a nuclear spin, and may therefore experience a hyperfine interaction described by
\begin{equation}\label{Eq: HFInt}
    \mathcal{H}_{hf} = \sum_{i=1}^{\tilde{N}} J_z^i \hat{z} \mathbb{A}_i \mathbf{I}^i,
\end{equation}
where $\tilde{N}$ is the total number of impurities, $\mathbf{I}^i$ ($\mathbf{J}^i$) is the $i$-th impurity nuclear (electron) spin operator and $\mathbb{A}_i$ is the corresponding hyperfine tensor. The interaction in Eq.~(\ref{Eq: HFInt}) typically leads to a complex energy structure which results in rich dynamics of the spin bath specially when including driving terms (see Supplementary Material, Section \ref{SM: Energy Structure}). In particular, on the one hand, the hyperfine energy splitting induces additional  frequencies that need to be resonantly targeted in order to achieve effective decoupling, thus requiring extra driving channels. On the other hand, in the case of LG driving, some hyperfine-suppressed flip-flop dynamics revive (see Supplementary Material, Section \ref{SM: Protocols}), worsening the method. Fortunately, our Hybrid-LG protocol can solve the latter issue by simultaneously applying resonant and off-resonant bath drivings, which yields improved central spin coherence.

To fully capture the decoherence dynamics of the central spin, many bath spins need to be considered. However, the exact simulation of large spin systems remains prohibitive due to its high dimensionality and therefore numerical approximation methods need to be used. We approximate the coherence of the central spin, $\mathcal{L}(2\tau) = 2\langle S_x (2\tau)\rangle$ using a variant of the well-known CCE method \cite{CCE-1,CCE-2}. In the standard CCE, the coherence is approximated by the product of all possible contributions of spin clusters $\mathcal{C}$ in the bath, truncated at some cluster size $M$, $\mathcal{L} \approx \mathcal{L}^{(M)}$, which sets the order of the approximation:
\begin{equation}
    \mathcal{L}^{(M)} = \prod_{\mathcal{C}|dim(\mathcal{C}) \leq M} \tilde{\mathcal{L}}_{\mathcal{C}}, \qquad \tilde{\mathcal{L}}_{\mathcal{C}} = \frac{\mathcal{L}_{\mathcal{C}}}{\prod_{\tilde{\mathcal{C}} \in \mathcal{C}} \tilde{\mathcal{L}}_{\tilde{\mathcal{C}}}}.
\end{equation}
Due to this construction, numerical instabilities may appear if two strongly coupled bath spin are separated into different clusters, producing unphysical results ($|\mathcal{L}| > 1$). To avoid this, it is standard to account for the interaction between bath spins in and outside a given cluster at mean-field level. That is, for a particular cluster, the surrounding spins are considered static along the direction of the external field, such that $\langle J_x \rangle = \langle J_y \rangle = 0$ and $\langle J_z \rangle = \pm 1/2$, the latter depending on the initial state. With this approximation, numerical instabilities are suppressed after averaging the coherence function over many initial states. However, when the bath is driven, numerical instabilities reemerge, as collective dynamics are no longer aligned with the external field, but along the set driving axis, rendering the previous assumption unsuitable. To account for this, we take the spin projections to be locked along the effective driving axis $P$, that is, $\langle J_P \rangle = \pm 1/2$ and $\langle J_{Q} \rangle = \langle J_{Q^\perp} \rangle = 0$, with $Q$ and $Q^\perp$ are the orthogonal directions to $P$. Therefore, the condition for static behavior is set by the driving, which can differ between subsets of bath spins with distinct energy splitting induced by Eq.~\eqref{Eq: HFInt}, complicating the mean-field treatment. On top of this, further subtleties arise between these subsets as their energy difference suppresses the flip-flop term in Eq.~\eqref{Eq: DipInt}, modifying the resulting mean-field contribution. These issues are addressed in our numerical framework where we first transform the pertinent interaction in Eq.~\eqref{Eq: DipInt} into the dressed basis defined by the drivings and then apply the mean-field approximation, for details see Supplementary Information, Section \ref{SM: NumSims}. Remarkably, we find that resurgent numerical instabilities are suppressed with mean field averaging, opening avenues for the study of decoupling protocols applied to baths featuring energetic splittings.

\begin{figure}[t]
    \centering
    \includegraphics[width=0.9\linewidth]{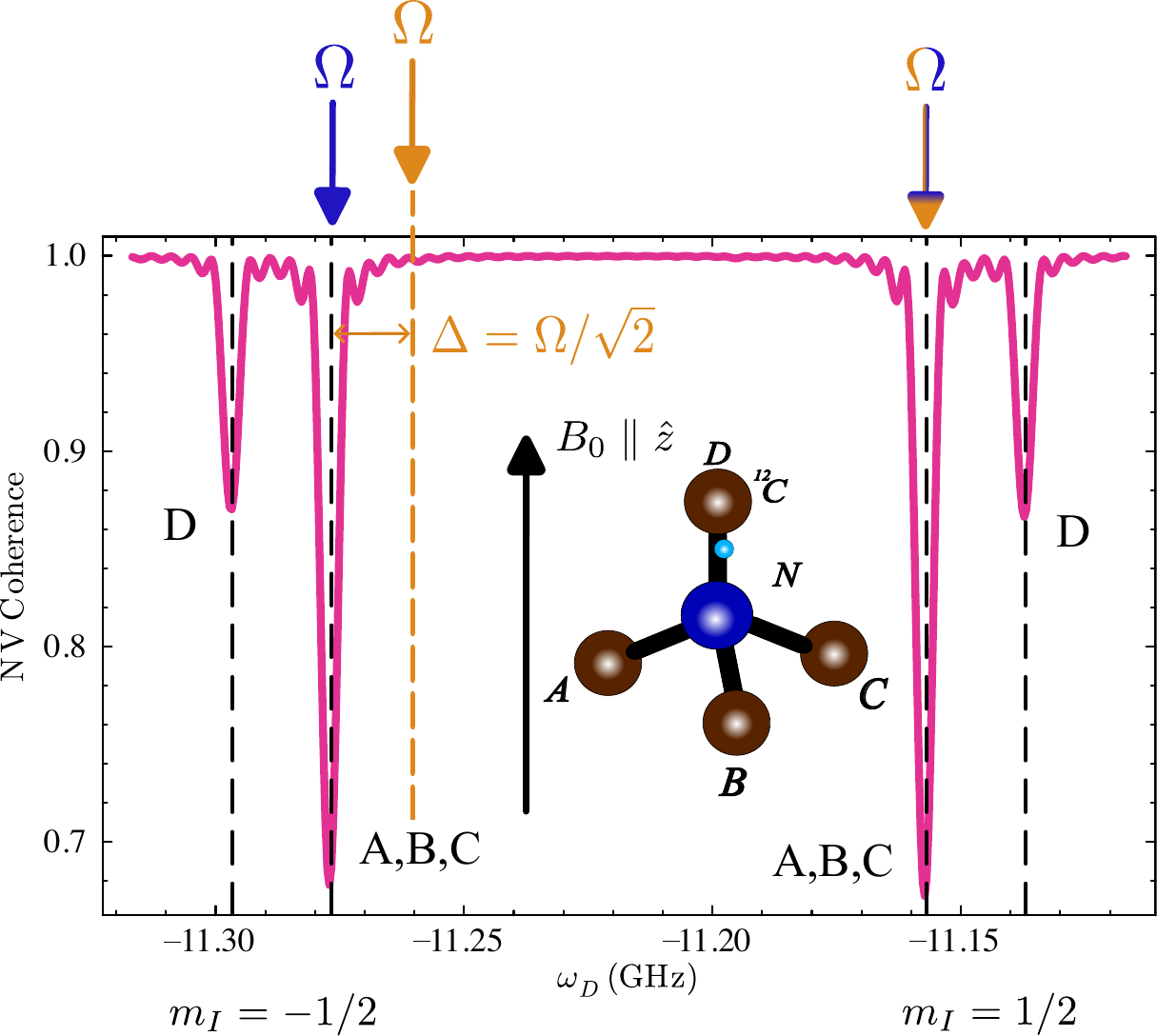}
    \caption{Simulated resonance structure of \ce{^15N}-P1 center bath in diamond. A hyperfine interaction produces different energy configurations depending on both the orientation of the P1 center in the lattice and the nitrogen spin state. The labels D (A, B, C) indicate orientations aligned (misaligned) with the external field $B_0$, here taken to be $B_0 = 0.4$ T. The target frequencies of the two-tone resonant (Hybrid-LG) bath drivings are indicated by blue (orange) arrows. Note that, in the Hybrid-LG protocol, one of these is detuned from the resonant frequency by an amount $\Delta = \Omega / \sqrt{2}$.}
    \label{F: P1B Driving}
\end{figure}

Implementation of the mean-field treatment in dense, strongly coupled spin baths requires extensive averaging, which is computationally expensive. For efficiency, we employ the partition CCE (pCCE) method, a modification of the standard CCE method introduced in \cite{SchatzlePCCE}. This consists on dividing the bath into partitions of strongly interacting spins and treating each partition as an entity in the CCE methodology. By doing so, one ensures that the strongest interactions among bath spins are always considered in the calculation of the coherence function, reducing the appearance of numerical instabilities as well as the number of internal averages required to achieve convergence. For more details on the numerical methods employed, see Supplementary Material, Section \ref{SM: NumSims}.

\begin{figure}[b]
    \centering
    \includegraphics[width=0.95\linewidth]{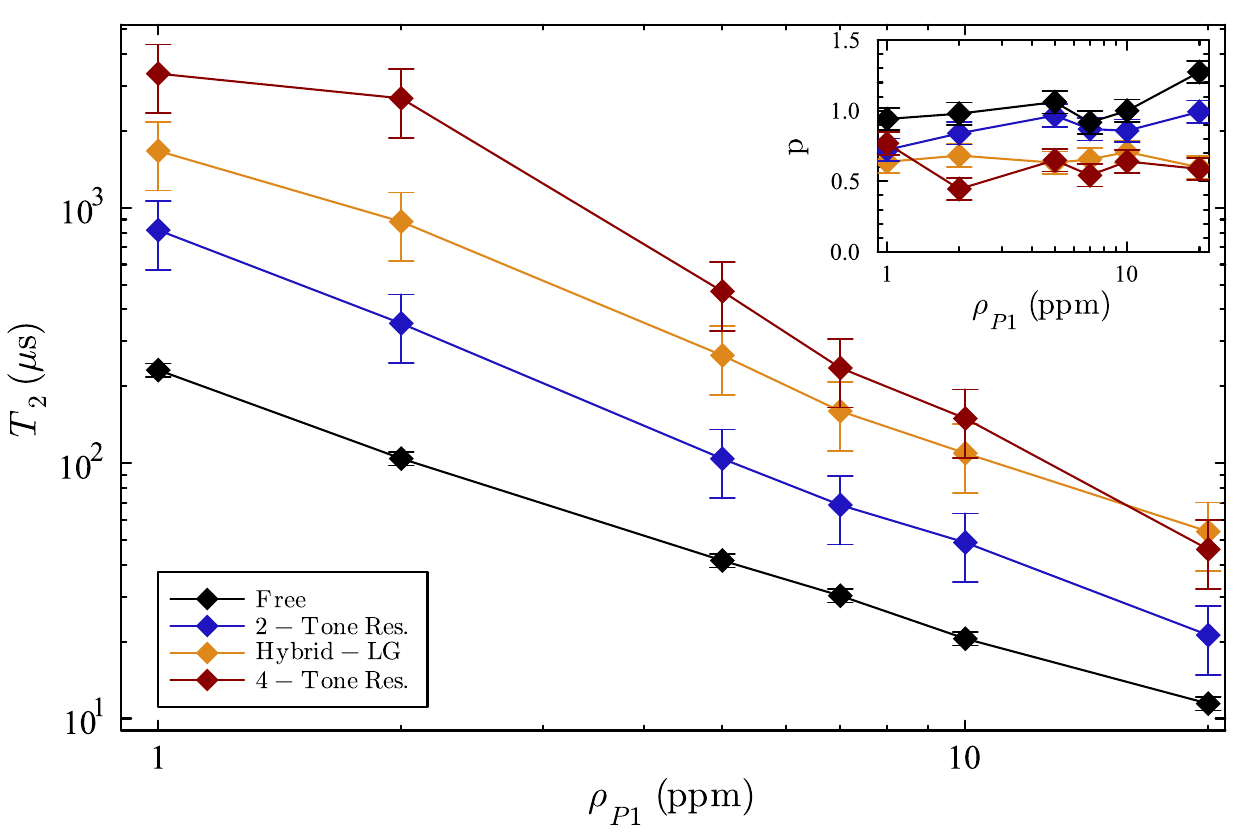}
    \caption{Hahn echo coherence time $T_2$ of an ensemble of NV centers as a function of the P1 impurity concentration $\rho_{P1}$ for different spin bath driving protocols, simulated with pCCE(2,4). All points are extracted from the coherence function, $\mathcal{L}(\tau)$, simulated as the average over 50 NV centers, each coupled to a unique bath of 180 randomly distributed P1 impurities. The sampled pulse spacing ($\tau$) of the Hahn echo was chosen to avoid resonance with the Rabi frequency ($\Omega$). In every protocol, each driving tone has $\Omega = 7 \,\mathrm{MHz}$. For no bath driving (black), the characteristic $T_2 \propto 1/\rho_{P1}$ scaling is observed. Under bath driving, a systematic two-fold increase in the coherence time is achieved using Hybrid-LG (orange) with respect to the standard resonant approach (blue). Furthermore, decoupling of Hybrid-LG is comparable to the four-tone resonant protocol (red). The inset shows the values of the stretched-exponential parameter $p$ for the same impurity concentrations. }
    \label{F: P1Bath}
\end{figure}

We showcase our model in a well-characterized solid-state platform based on diamond spin technologies: a sparse ensemble of nitrogen–vacancy (NV) centers coupled to the ubiquitous bath of substitutional nitrogen impurities, namely P1 centers or \ce{N^0_s}. Each P1 center consists of an electron spin coupled to a nuclear spin of either \ce{^14N} or \ce{^15N}, via the hyperfine interaction described in Eq.~\eqref{Eq: HFInt}. In many diamond samples the nitrogen content is purified to \ce{^15N}, which provides a long-lived nuclear memory enabling efficient repetitive readout of the NV center \cite{Arunkumar_RR}. Motivated by this, we focus on P1 centers containing \ce{^15N}, whose hyperfine structure yields four well-resolved resonances (see Fig.~\ref{F: P1B Driving}). Although the NV center is a spin-1 defect, we restrict attention to the two-level subspace ${|u\rangle,|d\rangle}={|0\rangle,|\pm1\rangle}$, for which $S_z=(\mathbb{1}\pm\sigma_z)/2$ with $\sigma_z$ the Pauli-$z$ matrix.

In the following, we present the ensemble coherence, obtained by averaging over 50 single-NV simulations in which each NV interacts with a bath of 180 \ce{^15N}-P1 spins at concentrations $\rho_{\mathrm{P1}} = 1$-$20$ ppm, each characterized by an independent random spatial configuration. Results are shown for various bath-driving protocols, including our Hybrid-LG approach. All simulations were performed using the high-performance computing facilities of the Donostia International Physics Center.\footnote{See the Donostia International Physics Center (DIPC) \href{https://dipc.ehu.eus/en/supercomputing-center}{webpage}.}
We find the ensemble coherence function to be well described by a second order pCCE with four P1 centers in each partition, i.e., a pCCE(2, 4) approximation. The results for different bath decoupling protocols are shown in Fig. \ref{F: P1Bath}, where the Hahn echo coherence $T_2$ time is plotted as a function of the P1 concentration $\rho_{P1}$ in a double-logarithmic scale. We obtain the coherence time by fitting the numerical data to a stretched-exponential function, $\mathcal{L}(2\tau) = e^{-(2\tau / T_2) ^p}$, which is expected to accurately describe the coherence decay of the central spins (for more details see the Supplementary Information, Section \ref{SM: Fitting}). The inset in Fig. \ref{F: P1Bath} shows the corresponding stretched-exponential parameter $p$ for the cases considered. In particular, we find $p\simeq 1$ for the case of no bath driving and $p < 1$ with driving, where smaller values are seen for schema with greater coherence protection. This is in agreement with previous studies  \cite{Bauch2020,Ghassemizadeh2024,SchatzlePCCE}.

Furthermore, our numerical results show that, in the case where the bath is not driven (in black), the coherence time scales as the inverse of the P1 concentration, $T_2 = A / \rho_{P1}$, as is to be expected \cite{Bauch2020}. We find the scaling factor to be $A = 217 \pm 13 \mathrm{\ \mu s \cdot ppm}$ for the \ce{^15N}-P1 bath, whereas previous numerical studies have found $A = 253 \pm 13 \mathrm{\ \mu s\cdot ppm}$ for a \ce{^14N}-P1 bath \cite{SchatzlePCCE}. This discrepancy arises from the fact that, in the latter case, the P1 centers show an additional resonant frequency which results in a larger suppression of flip-flop dynamics within the bath, thus leading to a longer coherence time. For the case in which the spin bath is resonantly driven (in blue), the ensemble coherence time is extended by a factor of three. This approach applies a two-tone resonant driving targeting the most populated bath energy states in Fig.~\ref{F: P1B Driving}, each corresponding to a Rabi frequency $\Omega = 7$ MHz. Further decoupling can be achieved either by addressing additional resonant frequencies of the bath and/or by increasing the Rabi frequency of each driving. To exemplify the former, we include simulations of the system under four tone resonant driving (in red). Although an improvement in coherence time is observed, with two extra tones this approach demands a four times extra power input than driving two tones (in blue). This may be impractical to implement experimentally and can cause significant heating that degrades ensemble coherence and/or enforces the reduction of the duty cycle \cite{knowles-bathDriving}.

Alternatively, our Hybrid-LG protocol does not rely on the addition of extra tones or power, but instead introduces a frequency detuning fulfilling the \textit{magic} condition to one of the two tones (in orange; see Fig. \ref{F: P1Bath}). In this scheme, the detuned tone suppresses a selection of the intra-bath dipolar interactions, and the resonant eliminates the remaining couplings between spins with different energy states; thereby preventing the revival of flip-flop dynamics and partially decoupling the bath from the central spin. Combined with Hahn echo refocusing, the resulting performance is captured in Fig.~\ref{F: P1Bath}. Remarkably, our protocol (in orange) exhibits a two-fold improvement in $T_2$ with respect to the two tone resonant driving scheme, while achieving similar values to those exhibited by the four tone driving approach, which requires a much greater power consumption. In light of these results, we conclude that the Hybrid-LG protocol offers an optimal balance between coherence protection and experimental practicality.

Moreover, we anticipate that our scheme can be improved further. For instance, a stroboscopic implementation of Hybrid-LG could be used, in which the targeted energy state for the magic driving is periodically switched. This approach could further reduce the total power input while preserving the decoupling performance. Besides, more sophisticated LG approaches, such as Frequency-Switch LG \cite{FSLG} or LG4 \cite{Halse_LG4} protocols, have demonstrated enhanced suppression of dipolar interactions as well as robustness against experimental errors, potentially yielding additional improvements. Finally, our framework could be extended to non-sparse qubit ensembles by incorporating more sophisticated MW control over the central spins \cite{AinitzeNVs}.

In this Letter we propose Hybrid-LG, a protocol to minimize dipolar interactions within a spin bath in solid-state systems via simultaneous application of resonant and off-resonant fields. To benchmark its performance, we compute the coherence time of an ensemble of NV centers in a dense \ce{^15N}-P1 bath and find a twofold increase relative to standard resonant bath driving. To accurately simulate the system, we develop a numerical framework based on the well-known CCE, which has proven to be an extremely useful tool for large spin system simulation. While these methods have been applied to the study of many solid-state platforms, including quantum dots, spin chains and bi- and three-dimensional crystal lattices \cite{CCE-1,CCE-2,SchatzlePCCE,ParkNature,Park2025NVDecoherence}, they have yet to incorporate spin bath driving. Here, we establish a mean-field theory to include driven bath dynamics, avoiding the unphysical divergences common in this context. We anticipate the developed framework to be applicable to the study many different solid-state systems.

\begin{acknowledgements}
We thank Philip Schäztle for useful discussions regarding the pCCE method. A. B. U. acknowledges the financial support of the IKUR STRATEGY IKUR-IKA-23/04. J. C. acknowledges the Ramón y Cajal (RYC2018-025197-I) research fellowship. This study is supported by the European Union’s Horizon Europe – The EU Research and Innovation Programme under grant agreement No 101135742 (”Quench”), the Spanish Government via the Nanoscale NMR and complex systems project PID2021-126694NB-C21, and the Basque Government grant IT1470-22. The authors acknowledge the technical and human support provided by the DIPC Supercomputing Center. 
\end{acknowledgements}

\bibliography{lib}


\newpage
\onecolumngrid
\include{SM}

\end{document}

%% file: SM.tex





\setcounter{section}{0}
\setcounter{figure}{0}
\renewcommand{\thefigure}{S\arabic{figure}}
\setcounter{table}{0}
\setcounter{equation}{0}

\begin{center}
    {\fontsize{18}{60}\bfseries\selectfont Supplementary Information} ~\\[0.5cm]
\end{center}

\renewcommand{\theequation}{S.\arabic{equation}}
\renewcommand{\thetable}{S.\arabic{table}}

\onecolumngrid
\restoreTOC
\tableofcontents

\section{The System}\label{SM: System}

In this work, we study a sparse ensemble of solid-state qubits where the inter-qubit couplings can be safely neglected. Each qubit --or central spin--, loses coherence due to its surrounding environment formed by a spin-1/2 bath (see Fig \ref{F: System} in the main text). The Hamiltonian describing such system is the following:
\begin{equation}\label{SMEq: SystemHamiltonian}
    \mathcal{H} = \mathcal{H}_{S} + \mathcal{H}_{B} + \mathcal{H}_{SB},
\end{equation}
where $\mathcal{H}_S$ ($\mathcal{H}_B$) is the central spin (bath) Hamiltonian and $\mathcal{H}_{SB}$ describes the interactions between them. In the following, we describe in detail each term in Eq. (\ref{SMEq: SystemHamiltonian}) and give the explicit expressions for the particular case of a nitrogen-vacancy (NV) center in diamond surrounded by a bath of substitutional nitrogen paramagnetic impurities (P1 centers). 

\subsection{Central Spin Hamiltonian: The NV Center} 

For an NV center acting as the central spin, the Hamiltonian is
\begin{equation}\label{SMEq: NV_H0}
    \mathcal{H}_S = \mathcal{H}_{NV} = DS_z^2 - \gamma_e B_0 S_z,
\end{equation}
where $D/2\pi = 2.88$ GHz is the so-called zero-field splitting, $\gamma_e/2\pi = -28.024$ GHz/T the electron gyromagnetic ratio and $\mathbf{S} = (S_x, S_y, S_z)$ the NV spin-1 operator. To lift the degeneracy of the $|\pm 1\rangle$ eigenstates, we employ an external bias magnetic field $\mathbf{B}_0$ along the NV axis, which we assume to be aligned with the $z$-direction. Additionally, a control driving $\mathcal{H}_c$ is applied to the NV center,
\begin{equation}\label{SMEq: Control}
    \mathcal{H}_c (t)= \sqrt{2}\Omega_{MW} \cos(\omega t + \phi) S_x,
\end{equation}
with $\Omega_{MW}$ and $\phi$ its Rabi frequency and phase. The frequency of the driving $\omega$ is tuned to match the $|0\rangle \to |\pm1\rangle$ transition, so that the NV can be treated as a spin-1/2 system. For clarity, we focus on the $|0\rangle \to |+1\rangle$ transition in the analytical development. Note that, then, $S_z = (\mathbb{1} + \sigma_z) / 2$, $\sigma_z$ being the third Pauli matrix. Moving to the rotating frame with respect to Eq. \eqref{SMEq: NV_H0}, and invoking the rotating wave approximation, the control Hamiltonian is rewritten as
\begin{equation}\label{SMEq: NV_Pulses}
    \mathcal{H}_c^I = \Omega_{MW}S_\phi,
\end{equation}
where $S_\phi = \sigma_\phi / 2$ and $\sigma_\phi = \cos\phi\sigma_x \mp \sin\phi\sigma_y$. The Hamiltonian (\ref{SMEq: NV_Pulses}) describes a pulse which induces population transfer between the two resonant NV energy levels; for instance, a pulse of duration $t_p = \pi/(2\Omega)$ (a $\pi$-pulse) induces population inversion between states $|0\rangle$ and $|+1\rangle$.

\subsection{Spin Bath Hamiltonian}\label{SM: Bath Hamiltonian}

The bath Hamiltonian $\mathcal{H}_{B}$ contains two terms: one describing the individual bath spins ($\mathcal{H}_{single}$) and another accounting for their mutual interactions ($\mathcal{H}_{BB}$), this is
\begin{equation}
	\mathcal{H}_{B}=\mathcal{H}_{single}+\mathcal{H}_{BB}.
\end{equation}
In the presence of an external magnetic field, the single spin term includes a Zeeman interaction that reads
\begin{equation}\label{SMEq: Zeeman}
    \mathcal{H}_{single}=\mathcal{H}_Z = \sum_{i=1}^N \omega_L^i J_z^i,
\end{equation}
where $\omega_L^i \equiv \gamma_i B_0$ is the Larmor frequency of the $i$-th bath spin, with $\gamma_i$ the gyromagnetic ratio, and $\mathbf{J}^i$ the corresponding spin operator. On the other hand, the bath spins interact with each other via dipole-dipole interactions. When such bath is composed of nuclear and electron spins, the intra-bath interactions are described by
\begin{equation}\label{SMEq: BathInt}
    \mathcal{H}_{BB} = \sum_{i=1}^{N_e} \sum_{j < i}^{N_e} \mathcal{H}^{ee}_{ij} + \sum_{i=1}^{N_n} \sum_{j < i}^{N_n} \mathcal{H}^{nn}_{ij} + \sum_{i=1}^{N_e} \sum_{j=1}^{N_n} \mathcal{H}^{en}_{ij},
\end{equation}
where $N_e$ ($N_n$) is the number of electron (nuclear) spins in the bath, $\mathcal{H}^{ee}_{ij}$ ($\mathcal{H}^{nn}_{ij}$) is the dipolar interaction between the electron (nuclear) spins and $\mathcal{H}^{en}_{ij}$ is the crossed dipolar interaction between them. Note that $N = N_e + N_n$.  Because the electron gyromagnetic ratio is much larger than the nuclear ($\gamma_e \sim 10^3 \gamma_n$), the last two terms in Eq. (\ref{SMEq: BathInt}) can be neglected if the bath contains electron spins. 

The expression for the dipolar interaction between two spins of any kind --electronic or nuclear-- reads
\begin{equation}\label{SMEq: DipInt}
    \mathcal{H}_{ij} = \frac{\gamma_i\gamma_j\mu_0\hbar}{4\pi r_{ij}^3}[\mathbf{J}^i \cdot \mathbf{J}^j - 3(\mathbf{J}^i\cdot\hat{r}_{ij})(\mathbf{J}^j\cdot\hat{r}_{ij})],
\end{equation}
with $\mathbf{r}_{ij} = r_{ij}\hat{r}_{ij}$ being the relative vector connecting both spins. If the spins $i$ and $j$ have similar Larmor frequency, such that $\abs{\omega_L^i -\omega_L^j}\ll \frac{\gamma_i\gamma_j\mu_0\hbar}{4\pi r_{ij}^3}$, in the rotating frame of the bath Zeeman energy $\mathcal{H}_Z$, the dipolar interaction is reduced to
\begin{equation}\label{SMEq: FlipFlop}
    \mathcal{H}_{ij}^I = C_{ij} \left[J_z^i J_z^j - \frac{1}{2}(J_x^i J_x^j + J_y^i J_y^j) \right],
\end{equation}
where the coupling constant is now
\begin{equation}
    C_{ij} = \frac{\gamma_i \gamma_j \mu_0 \hbar}{4\pi r_{ij}^3} (1 - 3\hat{z}_{ij}^2).
\end{equation}

For the case in which spins $i$ and $j$ have different Larmour frequencies, i.e.,  $\abs{\omega_L^i -\omega_L^j}\gg \frac{\gamma_i\gamma_j\mu_0\hbar}{4\pi r_{ij}^3}$,  in the rotating frame of the bath Zeeman energy $\mathcal{H}_Z$, the dipolar interaction is further reduced to
\begin{equation}\label{SMEq: NoFlipFlop}
    \mathcal{H}_{ij}^I = C_{ij} \left[J_z^i J_z^j \right],
\end{equation}
where the flip-flop dynamics are suppressed. This suppression is guaranteed for interactions between spins of different species, but it may also arise from other mechanisms that effectively introduce a detuning in the precession frequencies of two spins of the same species. For example, in crystal lattices, it is common to find paramagnetic defects composed of an electron spin coupled to a host nuclear spin, where the hyperfine interaction induces electronic subpopulations with different precession frequencies. That is, when the hyperfine coupling is present, the single term Hamiltonian of bath spin reads

\begin{equation}\label{SMEq: HFInt}
    \mathcal{H}_{\mathrm{single}}
    = \mathcal{H}_{Z} + \mathcal{H}_{\mathrm{hf}},
    \quad \text{where} \quad
    \mathcal{H}_{\mathrm{hf}}
    = \sum_{i=1}^{\tilde{N}} \mathbf{J}^{i} \cdot \mathbb{A}^{i} \cdot \mathbf{I}^{i}, 
\end{equation}
 $\tilde{N}$ being the total number of defects, $\mathbf{I}^i$ the $i$-th nuclear host spin operator and $\mathbb{A}^i$ the hyperfine tensor. Because the electron gyromagnetic ratio is typically larger than the hyperfine coupling, the hyperfine interaction in Eq. (\ref{SMEq: HFInt}) becomes, in the rotating frame with respect to the electron Zeeman energy,
\begin{equation}\label{SMEq: HFSplit}
    \mathcal{H}_{hf}^I = \sum_{i=1}^{\tilde{N}} J_z \hat{z}\mathbb{A}^i\cdot \mathbf{I}^i.
\end{equation}
Thus, the interaction in Eq. (\ref{SMEq: HFSplit}) results in a hyperfine energy shift of the electron spin that  changes the precession frequency, $\omega_L \to \omega_L + \Delta\omega \equiv \omega_t$, where $\Delta\omega$ depends on the hyperfine coupling amplitude and the state of the nucleus. Then, if the frequency detuning between a pair of spins is sufficiently large, their flip-flopping is suppressed, significantly altering the dynamics of the spin bath. A more detailed analytical development can be found in the next section.

\subsection{Bath Induced Decoherence}\label{SM: Decoherence}
The central spin is affected by the intra-bath dynamics through the interaction given by
\begin{equation}\label{Eq: SBHam}
    \mathcal{H}_{SB} = \sum_{i=1}^N \mathbf{S} \cdot \mathbb{A}_S^i \cdot \mathbf{J}^i,
\end{equation}
where $N$ is the number of bath spins, $\mathbf{S}$ is the central spin operator, $\mathbb{A}_S$ is the coupling tensor, and $\mathbf{J}^i$ are the operators of the electronic bath spins. Note that the interaction between the central spin and the nuclear bath spins is neglected due to small coupling strengths. In the rotating frame with respect to $\mathcal{H}_S$, the interaction in Eq. (\ref{Eq: SBHam}) under the secular approximation reads
\begin{equation}\label{SMEq: SecularInteractionI}
    \mathcal{H}_{SB} \simeq \sum_{i=1}^N S_z \mathbf{A}_S^i \cdot \mathbf{J}^i,
\end{equation}
which describes the pure dephasing dynamics of the central spin. Here, the interaction strength is
\begin{equation}
    \mathbf{A}_S^i = \frac{\gamma_S \gamma_i \mu_0 \hbar}{r_i^3}\left( \hat{z} - 3\frac{(\hat{z}\cdot\mathbf{r}_i)\cdot\mathbf{r}_i}{r_i^2} \right),
\end{equation}
with $\mathbf{r}_i = r_i\hat{r}_i$ the relative vector connecting the central spin with the $i$-th bath spin and $\gamma_S$ and $\gamma_i$ their corresponding gyromagnetic ratios. Now, a second secular approximation can be performed due to typically larger bath free energy compared to the coupling strength, so that the Hamiltonian (\ref{SMEq: SecularInteractionI}) is further reduced to
\begin{equation}\label{SMEq: SB}
    \mathcal{H}_{SB} = \sum_{i=1}^N S_z A_{S, z}^i J_z^i.
\end{equation}
The operators $J_z^i$ in this equation undergo oscillations generated by inhomogeneous flip-flop interactions in Eq. (\ref{SMEq: FlipFlop}) with other bath spins, giving rise to an important magnetic noise channel acting on the central spin. Particularly, when the bath spins are electronic, the intra-bath dipole-dipole couplings are strong, giving rise to high-frequency magnetic noise that rapidly dephases the central spin.

To illustrate the flip-flop dynamics of the electronic bath spins, consider a simple system consisting of a central spin and two crystal defects, each composed of an electron and a host nuclear spin. Neither the interaction with the central spin nor with the nuclei induces transitions of the electronic state. Thus, to simplify the calculation, we can assume that both the central spin and the nuclear spins are in a particular state $ |m_S, m_I^i, m_I^j\rangle$, where $m_S$ ($m_I^i$) is the magnetic number of the central spin (nucleus in the $i$-th crystal defect). Then, the Hamiltonian of the system is written as
\begin{equation}
    \mathcal{H}= (m_I^i A^i + m_S A_{S, z}^i) J_z^i + (m_I^j A^j + m_S A_{S, z}^j) J_z^j + C_{ij} [J_z^i J_z^j - 1/2(J_x^i J_x^j + J_y^i J_y^j)],
\end{equation}
where $\mathbf{J}^i$ is the spin operator of the electron in the $i$-th crystal defect.
As is usual, we may consider a pseudo-spin basis, $\{|\Uparrow\rangle, |\Downarrow\rangle\}$, where $|\Uparrow\rangle \equiv |\uparrow\downarrow\rangle$ and $|\Downarrow\rangle \equiv |\downarrow\uparrow\rangle$, such that
\begin{equation} 
    \mathcal{H}_\mathcal{B} = \left(\begin{array}{cc}
        \frac{A_\mathcal{B}}{2} - \frac{C_{ij}}{4} & -\frac{C_{ij}}{4} \\
        -\frac{C_{ij}}{4} & -\frac{A_\mathcal{B}}{2} - \frac{C_{ij}}{4}
    \end{array}\right) = \mathbf{h}_\mathcal{B}\cdot\tilde{\mathbf{I}} - \frac{C_{ij}}{4}\mathbb{1},
\end{equation}
with $A_\mathcal{B} \equiv (m_I^i A^i+ m_S A_{S, z}^i) - (m_I^j A^j + m_S A_{S, z}^j)$, $\mathbf{h}_\mathcal{B} \equiv (-C_{ij}/2, 0, A_\mathcal{B})$ and $\tilde{\mathbf{I}}$ a pseudo-spin 1/2 operator. Then, the flip rate between the two electron spins is found to be
\begin{equation}\label{SMEq: FlipRate}
    |\langle \Downarrow |\mathcal{U}_\mathcal{B} |\Uparrow\rangle| = \frac{C_{ij}/2A_\mathcal{B}}{\sqrt{1 + C_{ij}^2/4A_\mathcal{B}^2}}\sin\left[\frac{t}{4}\sqrt{4A_\mathcal{B}^2 + C_{ij}^2}\right].
\end{equation}
Defining $\omega_f \equiv \sqrt{4A_\mathcal{B}^2 + C_{ij}^2}$ and $\tan\theta \equiv \frac{C_{ij}}{2A_\mathcal{B}}$, Eq. (\ref{SMEq: FlipRate}) is rewritten as
\begin{equation}
    |\langle \Downarrow |\mathcal{U}_\mathcal{B} | \Uparrow\rangle| = \sin\theta\sin\left(\frac{\omega_f t}{4} \right),
\end{equation}
where the flip-flopping time scales is inversely proportional to the intra-bath coupling $C_{ij}$. An important conclusion follows from this result. When the detuning between the two electron spins is sufficiently large ($A_\mathcal{B} \gg C_{ij}$), the flip rate approaches zero, as $\theta \to 0$. Two main scenarios can lead to large $A_\mathcal{B}$: \textit{(i)} a large hyperfine detuning and \textit{(ii)} a significant difference in their coupling to the central spin, for example, due to substantially different spatial arrangement. It is important to note that, even if the flip-flopping magnitude is maximum when the two spins are perfectly degenerate ($A_\mathcal{B} = 0$), this will not be observed by the central as the transitions $| \Uparrow\rangle \to| \Downarrow\rangle$ commute with $J_z = J_z^1 + J_z^2$ which appear in Eq.~\eqref{SMEq: SB} due to symmetry. However, in reality, this symmetry is unlikely due to local homogeneities from other bath spins or central spin coupling and flip-flopping is present.

\section{Noise Reduction Protocols}\label{SM: Protocols}
\subsection{Detuned driving}
Moses Lee and Walter I. Goldburg showed that the continuous application of an off-resonant radio-frequency (RF) field may cancel, up to first order, the dipolar interaction between two homonuclear spins \cite{Lee-Goldburg}. In the same spirit, we consider a continuous bath driving acting on the bath in an attempt to suppress the noise affecting the central spin and extend its coherence time. For simplicity, we first assume no hyperfine interaction within the bath. In this case, the Hamiltonian describing the individual bath spin is
\begin{equation}
     \mathcal{H}_{single}(t)= \mathcal{H}_Z+  \mathcal{H}_D(t) \quad \text{where}\quad \mathcal{H}_D (t) = \sum_{i=1}^{N} 2\Omega \sin(\omega_{D}t + \alpha)J_x^i,
\end{equation}
where $\Omega$ is the Rabi frequency of the driving, $\omega_D$ the oscillation frequency and $\alpha$ its initial phase. In the rotating frame with respect to the bath Zeeman term in Eq. \eqref{SMEq: Zeeman},

\begin{equation}
    \mathcal{H}_D^I = \sum_{i=1}^N \Delta J_z^i + \Omega J_\alpha^i,
\end{equation}
where $\Delta = \omega_D - \omega_L$ is the frequency detuning of the driving and $J_\alpha = \sin\alpha J_x + \cos\alpha J_y$. Then the Hamiltonian describing the driven bath dynamics (in the rotating frame) reads
\begin{equation}
    \mathcal{H}_B^I = \sum_{i=1}^{N} \Delta J_z^i + \Omega J_\alpha^i + \sum_{j < i}^{N} C_{ij} \left[J_z^i J_z^j - \frac{1}{2}(J_x^i J_x^j + J_y^i J_y^j)\right],
\end{equation}
where we include the dipole-dipole interaction among bath spins, introduced in Eq. \eqref{SMEq: FlipFlop}. We can rewrite this coupling in the basis employed for the driving, that is $\{J_\alpha, J_{\alpha^\perp}, J_z\}$, with $J_{\alpha^\perp} = \cos\alpha J_x - \sin\alpha J_y$:
\begin{equation}
    \mathcal{H}_{ij}^{I} = C_{ij} \left[J_z^iJ_z^j - \frac{1}{2}(J_\alpha^iJ_\alpha^j + J_{\alpha^\perp}^iJ_{\alpha^\perp}^j)\right].
\end{equation}
Moving to another basis rotated around $\alpha^\perp$, in particular,
\begin{equation}\label{SMEq: Tilted}
    J_P = \cos\theta J_z + \sin\theta J_\alpha, \quad J_Q = \cos\theta J_\alpha - \sin\theta J_z, \quad J_{Q^\perp} = J_{\alpha^\perp},
\end{equation}
where 
\begin{equation}
    \cos\theta = \frac{\Delta}{\sqrt{\Delta^2 + \Omega^2}} \qquad \text{and} \qquad \sin\theta = \frac{\Omega}{\sqrt{\Delta^2 + \Omega^2}},
\end{equation}
the bath driving Hamiltonian can be simplified to
\begin{equation}\label{SMEq: EffD}
    \mathcal{H}_D^I = \sum_{i=1}^N \bar{\Omega} J_P^i \quad \text{where} \quad \bar{\Omega}=\sqrt{\Delta^2+\Omega^2}.
\end{equation}
In this basis, the intra-bath dipole-dipole interaction reads
\begin{align}\label{SME: IntHam2}
    \mathcal{H}_{ij}^I & = C_{ij} \left\{\cos^2\theta J_P^i J_P^j + \sin^2\theta J_Q^i J_Q^j - \sin\theta\cos\theta (J_P^i J_Q^j + J_Q^i J_Q^j)\right.
    \notag \\
    &\left. - \frac{1}{2} \left[\cos^2\theta J_Q^i J_Q^j + \sin^2\theta J_P^i J_P^j + \sin\theta\cos\theta (J_P^i J_Q^j + J_Q^i J_P^j) + J_{Q^\perp}^i J_{Q^\perp}^j\right]\right\}.
\end{align}
In a further interaction picture defined with respect to the bath driving term in Eq. \eqref{SMEq: EffD}, the dipole-dipole interaction can be reduced using the rotating-wave approximation, as long as $\bar{\Omega}\gg C_{ij}$. This yields
\begin{equation}\label{SMEq: IntHam}
    \mathcal{H}_{ij}^{II} = C_{ij} \left(\cos^2\theta - \frac{1}{2}\sin^2\theta\right) \left\{ J_P^i J_P^j  - (J_Q^i J_Q^j + J_{Q^\perp}^i J_{Q^\perp}^j)\right\}.
\end{equation}
Remarkably, the dipole-dipole interaction completely disappears when the so-called Lee-Goldburg (LG) condition --or  \textit{magic} angle condition-- is satisfied, i.e., when $\cos\theta = \pm 1 / \sqrt{3}$, or, equivalently, $\Delta = \pm \Omega/\sqrt{2}$. That is, when a strong enough driving is applied fulfilling the former condition, the flip-flop interaction of the bath spins is suppressed. Consequently, the interaction between the bath and the central spin remains time independent, and therefore can be easily refocused by the spin echo sequence. Besides, it is interesting to note, that this detuned continuos driving, not only cancels the intra-bath interactions, but also effectively reduces the coupling of the bath to the central spin in Eq. \eqref{SMEq: SB} to
\begin{equation}
    \mathcal{H}_{SB}^{II} = \sum_{i=1}^{N} \cos{\theta}\;A_{S, z}^i S_z J_P^i.
\end{equation}
Also, it is important to mention that within the bath there might be strong couplings that cannot be eliminated due to limited driving power, resulting in the eventual decay of the central qubit coherence, even under LG off-resonant bath driving.

\subsection{Resonant driving}
Previous experiments have used a resonant bath driving scheme to improve the central spin coherence \cite{knowles-bathDriving,Bauch2018,Barry2024}. In this scenario, i.e., $\Delta=0$, the driving field is orthogonal to $J^i_z$; thus the coupling of the bath spins to the central spin in Eq.~\eqref{SMEq: SB} is suppressed (again, this elimination remains imperfect due to the limited available driving power). On the other hand, the dipole-dipole interaction in Eq. \eqref{SME: IntHam2} yields
\begin{equation}
    \mathcal{H}_{ij}^{II} = -\frac{C_{ij}}{2} \left[J_\alpha^i J_\alpha^j - \frac{1}{2}(J_{\alpha^\perp}^i J_{\alpha^\perp}^j + J_z^i J_z^j) \right].
\end{equation}
That is, the flip-flop interaction persists.

\subsection{Hybrid driving}
When the bath spins exhibit hyperfine coupling, the dynamics differ. As described in the previous section, this  coupling induces shifts in the precession frequencies, leading to bath spin subsets with distinct resonance energies. This phenomena, on the one hand, results in the natural suppression of the flip-flop interaction among bath spin subsets leading to ZZ interaction as in Eq.~\eqref{SMEq: NoFlipFlop}; and on the other hand, increases the number of driving tones required to address the bath.

Furthermore, an inappropriate choice of driving scheme can lead to degraded performance --particularly when LG driving is applied simultaneously to different spin subsets. To illustrate this, consider the following Hamiltonian, which describes a system consisting of two bath spins (labelled $1$ and $2$) with distinct hyperfine shifts,
\begin{equation}
    \mathcal{H}_B^{I} = \sum_{i=1,2} \big( \Delta_i J_z^i + \Omega_i J_\alpha^i \big)+ C_{12} J_z^1 J_z^2.
\end{equation}
Note that the flip-flop interaction between these two spins is suppressed. Moving to the set of tilted basis in Eq. \eqref{SMEq: Tilted} defined by the particular choice of Rabi frequencies and detunings for each bath spin, leads to
\begin{equation}
    \mathcal{H}_B^{I} = \sum_{i=1,2}^{N} \bar{\Omega}_i J_P^i +  C_{12} \left\{ \cos\theta_1\cos\theta_2 J_P^1 J_P^2 + \sin\theta_1\sin\theta_2 J_Q^1 J_Q^2 - \sin\theta_1\cos\theta_2 J_Q^1 J_P^2 - \cos\theta_1\sin\theta_2 J_P^1 J_Q^2 \right\},
\end{equation}
where $\bar{\Omega}_i$ and $P$ can differ from one spin to another. Then, in the rotating frame of the effective Rabi frequencies, we find
\begin{align}\label{SMEq: IntHamII}
    \mathcal{H}_B^{II} & \approx C_{12} \left\{J_P^1 J_P^2 \cos\theta_1\cos\theta_2   \right. \notag \\
    & + \frac{\sin\theta_1\sin\theta_2}{2}\cos((\bar{\Omega}_1 - \bar{\Omega}_2)t)(J_Q^1 J_Q^2 + J_{Q^\perp}^1 J_{Q^\perp}^2) - \frac{\sin\theta_1\sin\theta_2}{2}\sin((\bar{\Omega}_1 - \bar{\Omega}_2)t)(J_{Q^\perp}^1 J_Q^2 - J_Q^1 J_{Q^\perp}^2) \notag \\
    & + \frac{\sin\theta_1\sin\theta_2}{2}\cos((\bar{\Omega}_1 + \bar{\Omega}_2)t)(J_Q^1 J_Q^2 - J_{Q^\perp}^1 J_{Q^\perp}^2) - \frac{\sin\theta_1\sin\theta_2}{2}\sin((\bar{\Omega}_1 + \bar{\Omega}_2)t)(J_{Q^\perp}^1 J_Q^2 + J_Q^1 J_{Q^\perp}^2) \notag \\
    & \left. -\cos(\bar{\Omega}_1 t) \sin\theta_1\cos\theta_2 J_Q^1 J_P^2 - \cos(\bar{\Omega}_2t) \cos\theta_1\sin\theta_2 J_P^1 J_Q^2 \right. \notag \\
    & \left. +\sin(\bar{\Omega}_1 t) \sin\theta_1\cos\theta_2 J_{Q^\perp}^1 J_P^2 + \sin(\bar{\Omega}_2 t) \cos\theta_1\sin\theta_2 J_P^1 J_{Q^\perp}^2\right\}.
\end{align}
We now assume that the effective Rabi frequency is much larger than the dipole-dipole coupling strength, or $C_{12} \ll \bar{\Omega}_{1/2}$, but relax the condition between the two unique frequencies $|\bar{\Omega}_1 - \bar{\Omega}_2|$, then this Hamiltonian (\ref{SMEq: IntHamII}) simplifies to 
\begin{align}\label{SMEq: IntHamIIRed}
    \mathcal{H}_B^{II} = & C_{12} \left\{\cos\theta_1\cos\theta_2 J_P^1 J_P^2 \right. \notag \\
    & \left. + \frac{\sin\theta_1\sin\theta_2}{2}\left[\cos((\bar{\Omega}_1 - \bar{\Omega}_2)t)(J_Q^1 J_Q^2 + J_{Q^\perp}^1 J_{Q^\perp}^2) - \sin((\bar{\Omega}_1 - \bar{\Omega}_2)t)(J_{Q^\perp}^1 J_Q^2 - J_Q^1 J_{Q^\perp}^2)\right]\right\}.
\end{align}
Importantly, if both spins are driven satisfying the same LG condition, the Hamiltonian (\ref{SMEq: IntHamII}) becomes
\begin{equation}
    \mathcal{H}_B^{II} = \frac{C_{12}}{3} \left[ J_P^1 J_P^2 + J_Q^1 J_Q^2 + J_{Q^\perp}^1 J_{Q^\perp}^2\right] = \frac{C_{12}}{3}\mathbf{J}^1\cdot\mathbf{J}^2
\end{equation}
If the two spins have the same effective Rabi, this scalar coupling between the two spins commutes with the collective driving terms such that it does not invoke flip-flopping. However, in realistic systems where each spin may be detuned due to interactions with the NV or other nearby bath spins, the symmetry of the scalar coupling is broken and the spins will flip-flop in a direction perpendicular to the driving. A more damaging effect can be seen if the spins match the magic angle condition with opposite signs, such that $\cos\theta_1 = 1/\sqrt{3}$ and $\cos\theta_2 = -1/\sqrt{3}$. Then, the interaction becomes
\begin{equation}
    \mathcal{H}_B^{II} = -\frac{C_{12}}{3} \left[ J_P^1 J_P^2 - J_Q^1 J_Q^2 - J_{Q^\perp}^1 J_{Q^\perp}^2\right],
\end{equation}
an antisymmetric interaction. Now, even without detuning, flip-flopping will emerge between spins with different hyperfine shifts. Hence, if the detuning is not chosen correctly, the use of LG driving will not result in increased coherence time, because even if the dipole-dipole interactions between pairs corresponding to the same spin subset are suppressed (see Eq.~\eqref{SMEq: IntHam}), flip-flop dynamics are induced for other pairs.

By observing the Hamiltonian \eqref{SMEq: IntHamIIRed}, there is a more robust condition which suppresses flip-flopping. If the effective Rabi frequencies are different, such that $|\bar{\Omega}_1 - \bar{\Omega}_2| \gg C_{12}$, then the terms invoking flip-flopping will be suppressed and the interaction returns to a simple form $\propto J_P^1 J_P^2$. Hence, as an alternative, in the main text we propose \textit{Hybrid-LG}, an approach in which a subset of the bath spins are driven resonantly, while others are driven magically, so that the remaining interaction between them is cancelled. This requires $|\bar{\Omega}_1 - \bar{\Omega}_2| \gg C_{12}$, however for strong enough driving, this is satisfied. As can be seen in the main text, the coherence of the central spin is better protected with \textit{Hybrid-LG} than when the bath is driven resonantly. Of course, one could also applying LG driving with different Rabi frequencies, however we find that a hybrid approach yields better protection. 

\section{The P1 Center}\label{SM: P1 Center}  

\begin{figure}[ht!]
    \centering
    \includegraphics[width=0.3\textwidth]{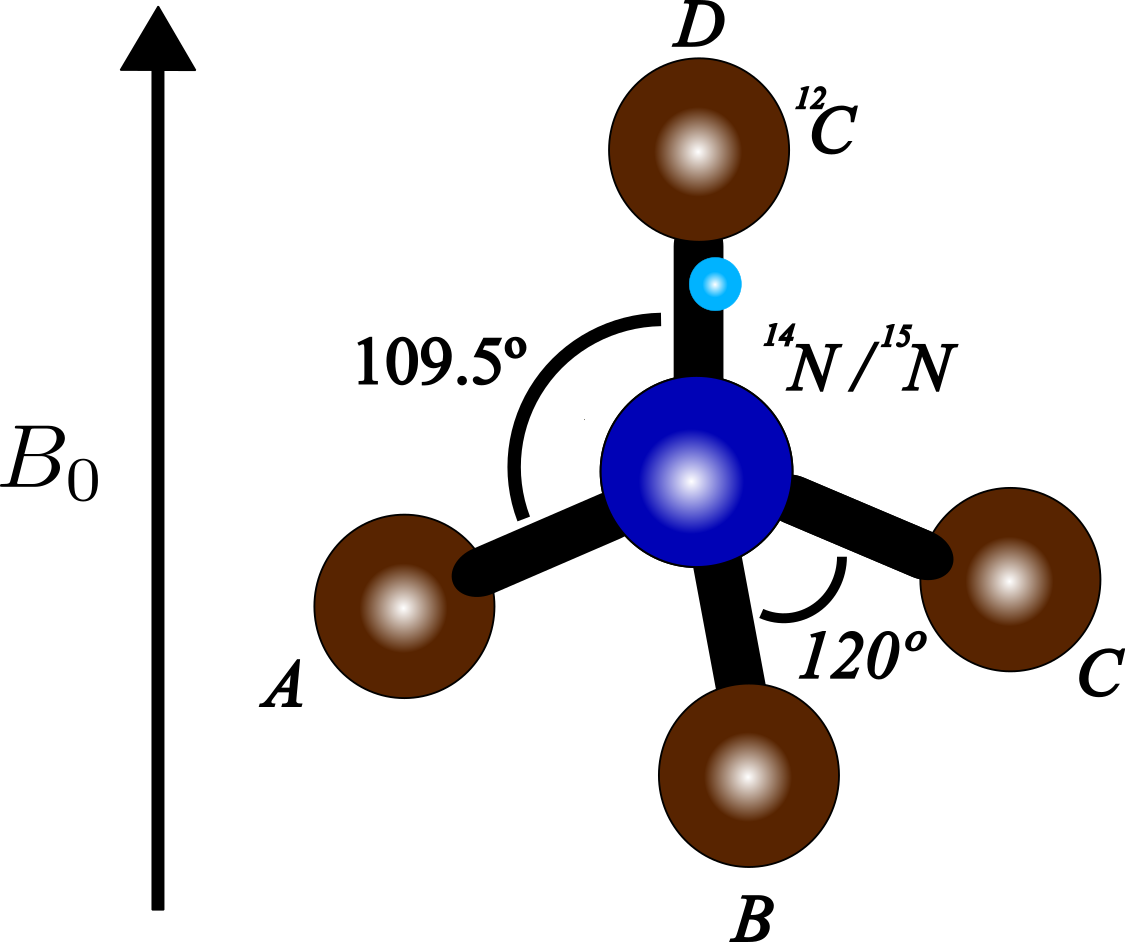}
    \caption{The P1 center in an external magnetic field. Carbon (\ce{^12C}) atoms are depicted in brown, the nitrogen (\ce{^14N}/\ce{^15N}) atom in blue and the electron in light blue. The labels A, B, C and D refer to the possible orientations of the center. The case shown corresponds to an \textit{on-axis} P1 center (labeled D).}
    \label{P1center}
\end{figure}

The formation of NV centers in diamond requires the presence of nitrogen nuclei in the lattice, which couple to a vacancy to form the NV center. However, not every nitrogen nuclei converts to NV, and thus residual paramagnetic impurities, which are called \ce{N^0_s} or P1 centers, emerge in the lattice. These are formed by a nitrogen nucleus that substitutes a carbon nucleus and a spare electron and are a source of electron magnetic noise that dephases the NV center, as we will see. Fig. \ref{P1center} depicts the P1 center and its four carbon neighbours.

The Hamiltonian describing the $i$-th P1 center is \cite{Bartling2025}
\begin{equation}\label{SMEq: P1 Ham}
    \mathcal{H}_{P1}^i = \gamma_e B_0 J_z^i + \gamma_N B_0 I_z^i + \mathbf{J}^i\cdot\mathbb{A}^i \cdot \mathbf{I}^i + \mathbf{I}^i\cdot \mathbb{P}^i \cdot \mathbf{I}^i,
\end{equation}
where $\mathbf{J}$ is the spin-1/2 operator of the electron, $\mathbf{I}$ is the nuclear spin operator (spin-1 for the \ce{^14N} isotope and spin-1/2 for \ce{^15N}), $B_0$ is the bias magnetic field and $\gamma_e$ ($\gamma_N)$ is the electron (nuclear) gyromagnetic ratio ($\gamma_{\ce{^14N}}/2\pi = 3.077$ MHz/T and $\gamma_{\ce{^15N}}/2\pi = -4.316$ MHz/T). The last two terms in Eq. (\ref{SMEq: P1 Ham}) are the hyperfine and quadrupolar Hamiltonians, with $\mathbb{A}$ and $\mathbb{P}$ the hyperfine and quadrupole tensors (note that the latter disappears for spin-1/2 nuclei, such as the \ce{^15N} isotope). 

 Besides, we consider room temperature conditions, where the initial state of each P1 electron spin is approximated as the mixed state,
\begin{equation}
    \rho_{P1} = \frac{1}{2}\left(\begin{array}{cc}
        1 & 0 \\
        0 & 1
    \end{array}\right).
\end{equation}

\subsection{Energy Structure of the P1 Center}\label{SM: Energy Structure}

The tetrahedral site symmetry ($T_d$) of diamond is reduced to trigonal ($C_{3v}$) for the P1 center due to the elongation of one of the four equivalent N-C bonds which localizes the electron on an anti-bonding orbital along this bond. The spare electron of the P1 center can occupy any of the four bonds with the neighbouring C nuclei, leading to four possible orientations of the P1 center in the diamond lattice, all equivalent in energy in the absence of strain or external interactions. Although the P1 center undergoes a reorientation process with a rate of $\sim$ kHz, for the timescales considered in this work ($\mathrm{\mu s}$), it can be considered static.

When placed in an external magnetic field aligned with any of the four orientations, a preferential orientation along the direction parallel to the field is established. This direction sets the \textit{principal} axis of the P1 center and the corresponding orientation is labeled $D$ or \textit{on-axis}. The remaining three orientations are labeled \textit{A, B} and \textit{C} or \textit{off-axis}. The hyperfine tensor for the P1 center depends on its orientation in the lattice, where \textit{on-axis} P1 centers exhibit larger hyperfine couplings, as they experience the full magnitude of the external field. The hyperfine tensor $\mathbb{A}$ can be written for any orientation as \cite{DegenDarkSpins}
\begin{equation}
    \mathbb{A} = R^T(\alpha, \beta, \gamma) \mathbb{A}_{diag} R(\alpha, \beta, \gamma),
\end{equation}
where $R(\alpha, \beta, \gamma)$ is the SO(3) rotation operator in Euler angles and the diagonal coupling $\mathbb{A}_{diag}$, which corresponds to the \textit{principal} reads, for different nitrogen isotopes,
\begin{align}
    \mathbb{A}_{diag}^{\ce{^14N}} & = (81.312, 81.312, 114.0264) \ \mathrm{ MHz}, \\
    \mathbb{A}_{diag}^{\ce{^15N}} & = (113.83, 113.83, 159.7) \ \mathrm{ MHz}.
\end{align}
Similarly, the quadrupole tensor can also be written, for any axis, as
\begin{equation}
    \mathbb{P} = R^T(\alpha, \beta, \gamma) \mathbb{P}_{diag} R(\alpha, \beta, \gamma),
\end{equation}
where
\begin{equation}
    \mathbb{P}_{diag} = (P_\perp, P_\perp, P_\parallel).
\end{equation}
Here, $P_\parallel = -3.977$ MHz. To our knowledge, no experimental value for $P_\perp$ has been reported, although Park \textit{et al.} \cite{ParkNature} have performed theoretical calculations for this and the hyperfine couplings via Density Functional Theory. Note that this value is for the \ce{^14N} isotope, as \ce{^15N} does not have a quadrupole moment, although, we find in our simulations that the quadrupolar interaction, as well as the nuclear Zeeman term, can be safely neglected.

Because of the axial symmetry of the P1 center in its \textit{principal} axis ($A_x = A_y, P_x = P_y$), we can set $\gamma = 0$ without loss of generality, so that the general rotation matrix
\begin{equation}
    R(\alpha, \beta, \gamma) = \left( \begin{array}{ccc}
        \cos\gamma \cos\beta \cos\alpha - \sin\gamma \sin\alpha & \cos\gamma \cos \beta \sin\alpha + \sin\gamma \cos\alpha & -\cos\gamma \sin\beta \\
        -\sin\gamma \cos\beta \cos\alpha - \cos\gamma \sin\alpha & -\sin\gamma \cos\beta \sin\alpha + \cos\gamma \cos\alpha & \sin\gamma \sin\beta \\
        \sin\beta \cos\alpha & \sin\beta \sin\alpha & \cos\beta
    \end{array}\right)
\end{equation}
simplifies to 
\begin{equation}
    R(\alpha, \beta) = \left( \begin{array}{ccc}
        \cos\beta \cos\alpha & \cos \beta \sin\alpha & -\sin\beta \\
        -\sin\alpha & \cos\alpha & 0 \\
        \sin\beta \cos\alpha & \sin\beta \sin\alpha & \cos\beta
    \end{array}\right).
\end{equation}
 Moreover, because the P1 center is symmetric in the \textit{x}-\textit{y} plane, the rotation matrix can be further simplified to polar angle rotations, $R(\alpha, \beta) = R_y(\beta)$, where $\beta = 0$ for the \textit{on-axis} orientation and $\beta = 109.5^\circ$ for the \textit{off-axis}. Then, defining $\mathbb{A}_{diag} \equiv (A_\perp, A_\perp, A_\parallel)$, the hyperfine term can be written as
 \begin{align}
    \mathcal{H}_{hf} & = \mathbf{J} \cdot \mathbb{A} \cdot \mathbf{I} = \mathbf{J} \cdot R_y^T(\beta) \mathbb{A}_{diag} R_y(\beta) \cdot \mathbf{I} \nonumber \\
    & = [A_+ + A_- \cos2\beta]J_zI_z - A_-\sin2\beta J_zI_x - A_-\sin2\beta J_x I_z + [A_+ - A_- \cos2\beta] J_x I_x \\
    &+ (A_+ - A_-) J_y I_y, \nonumber
 \end{align}
 where $A_\pm \equiv \dfrac{A_\parallel \pm A_\perp}{2}$. To find the energy shifts due to the hyperfine coupling, we can apply a secular approximation such that we consider only the terms that depend on $J_z$ and diagonalize them in the nitrogen subspace. For a \ce{^14N} host with spin 1, the hyperfine eigenenergies are
 \begin{equation}\label{SMEq: 14NHf}
    E_0 = 0, \quad E_\pm = \pm \sqrt{A_+^2 + A_-^2 + 2A_+A_-\cos2\beta},
 \end{equation}
and for a \ce{^15N} host with spin 1/2, we find
\begin{equation}\label{SMEq: 15NHf}
    E_{\pm 1/2} = \pm \frac{1}{2}\sqrt{A_+^2 + A_-^2 + 2A_+A_-\cos2\beta}.
\end{equation}
As a side note, using directly the general rotation matrix $R(\alpha, \beta, \gamma)$ to write the hyperfine interaction leads to the same hyperfine eigenenergies shown above, as expected. These shifts (which are given with respect to the electron Zeeman energy) correspond to different nitrogen isotopes and P1 orientations and are shown in Table \ref{T: hfEnergies}. For a \ce{^14N}-P1 center, there are five energy levels with the following populations \footnote{To obtain these, we assume that \textit{i)} the nitrogen is not polarized and \textit{ii)} every axis is equally likely. Should these assumptions not hold, then the populations of these energy levels would change.}: for off-axis centers with $m_I = \pm 1$, the population is 3/12 in both cases; for on-axis centers with $m_I = \pm 1$, the population is 1/12; for centers with $m_I = 0$, the population is $4/12$  because the hyperfine interaction disappears. Similarly, in the case of a \ce{^15N}-P1 center, there are four energy branches with the following populations: for off-axis centers with $m_I = \pm 1/2$, the population is 3/8 in both cases and for on-axis centers with $m_I = \pm 1/2$, the population is 1/8. Note that the nitrogen magnetic number is given in a dressed basis.

\begin{table}[ht!]
    \centering
    \caption{Hyperfine energy shifts for the different JT axes and nitrogen isotopes.}
    \begin{tabular}{llclc llcl}
        \hline\hline
        \multicolumn{4}{c}{\ce{^14N}} & & \multicolumn{4}{c}{\ce{^15N}} \\
        \hline
        \multirow{2}{*}{$\beta = 0$} 
            & $E_0$   & $=$ & $0$ MHz   & &
        \multirow{2}{*}{$\beta = 0$} 
            & $E_{\pm 1/2}$ & $=$ & $\pm 79.85$ MHz \\
        & $E_\pm$ & $=$ & $\pm 114.03$ MHz & & & & & \\
        \multirow{2}{*}{$\beta = 109.5^\circ$} 
            & $E_0$   & $=$ & $0$ MHz   & &
        \multirow{2}{*}{$\beta = 109.5^\circ$} 
            & $E_{\pm 1/2}$ & $=$ & $\pm 59.91$ MHz \\
        & $E_\pm$ & $=$ & $\pm 85.58$ MHz & & & & & \\
        \hline\hline
    \end{tabular}
    \label{T: hfEnergies}
\end{table}

\begin{figure}[t]

\subfloat[\label{F: DEER.a}]{%
  \includegraphics[width=0.45\columnwidth]{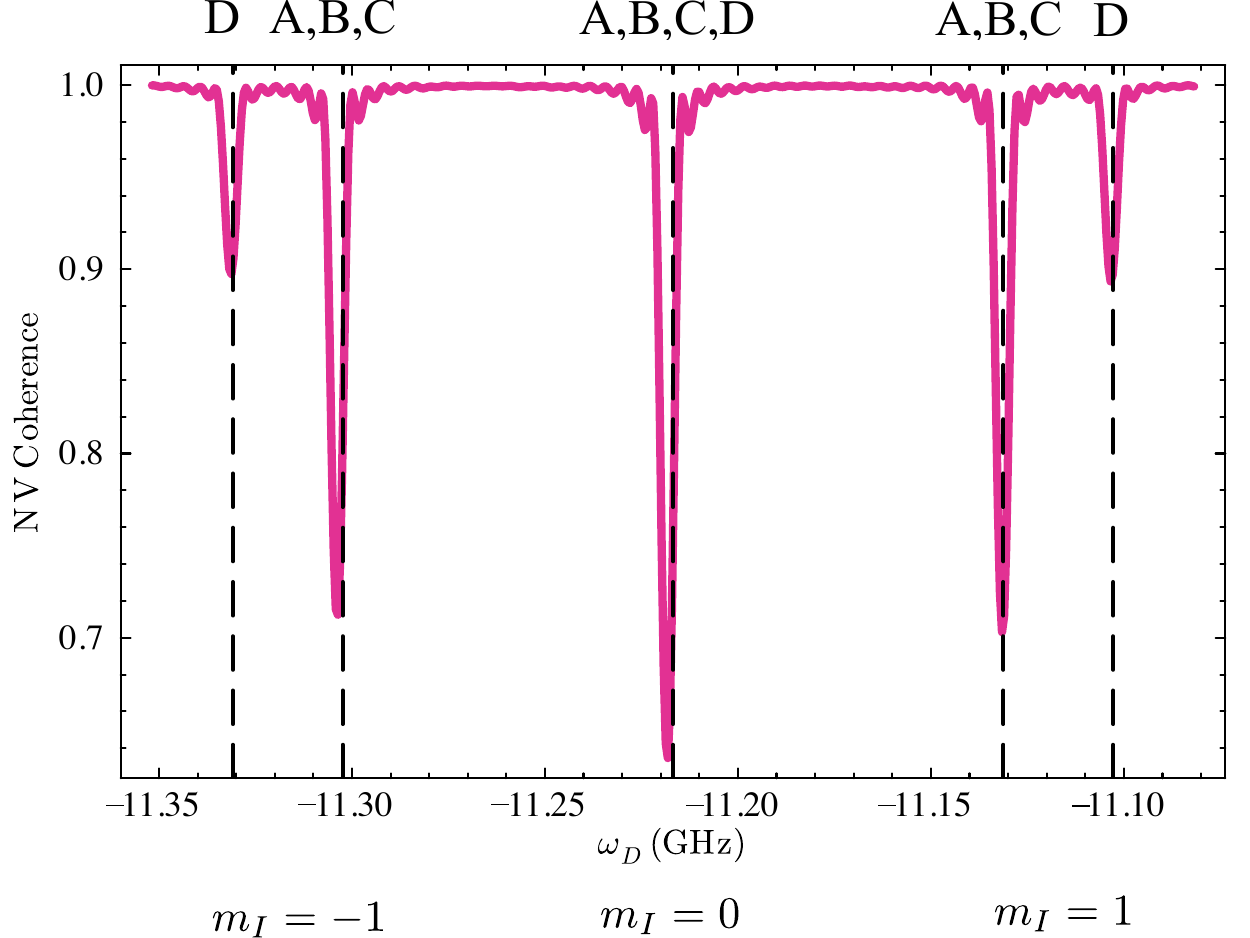}%
}\hfill%
\subfloat[\label{F: DEER.b}]{%
  \includegraphics[width=0.45\columnwidth]{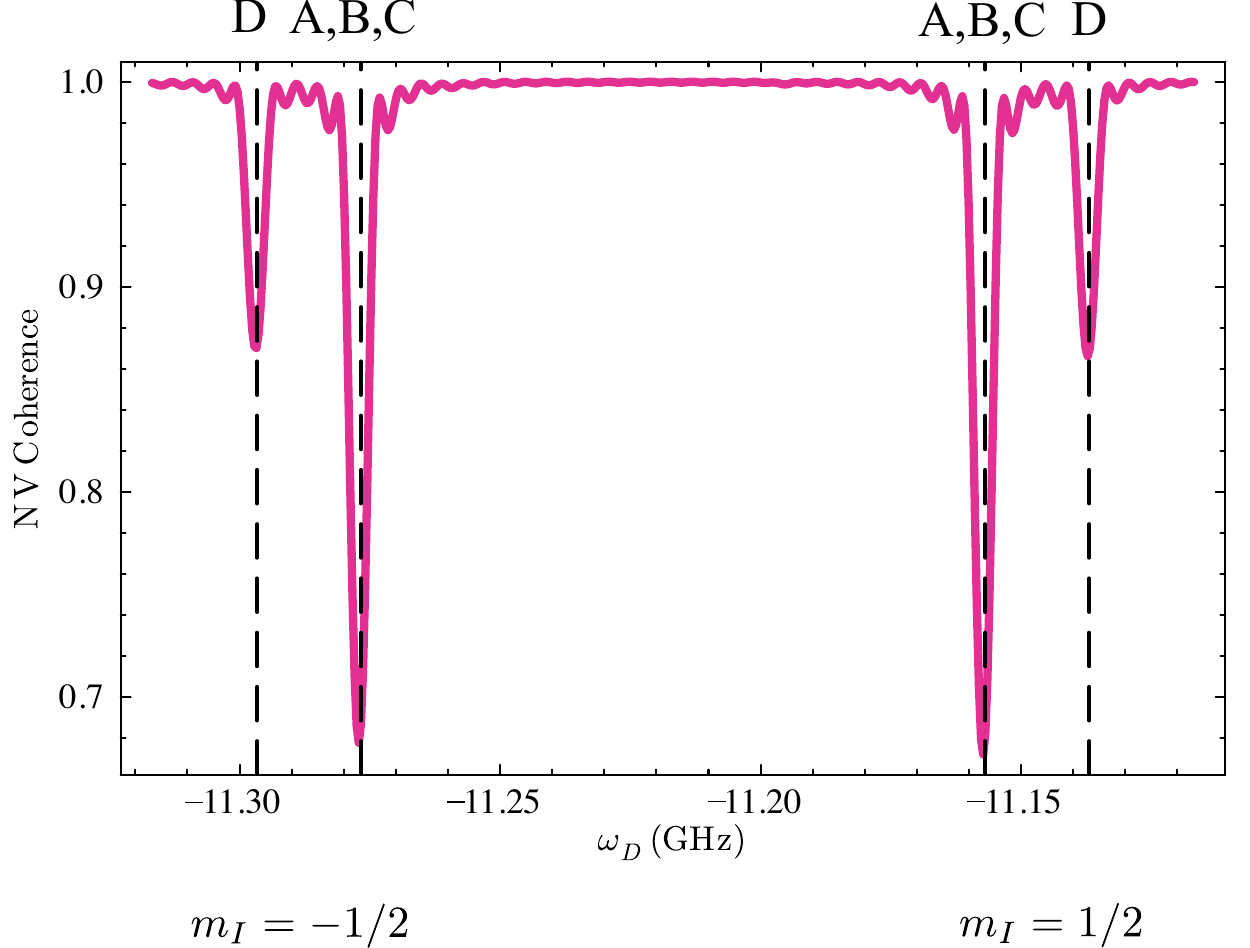}%
}
\caption{Simulated DEER Spectrum of a P1 bath formed with (a) \ce{^14N} and (b) \ce{^15N}. Dashed lines indicate the theoretical prediction of Eq. (\ref{SMEq: 14NHf}) and (\ref{SMEq: 15NHf}). The labeling of each resonance peak (top) indicates the corresponding orientation of the center, while the labels in the bottom indicate the nitrogen dressed state.} \label{F: DEER}
\end{figure}

Then, the hyperfine Hamiltonian in the nitrogen dressed basis reads
\begin{equation}
    \mathcal{H}_{hf}^i = \sum_{m_I^i} E_{m_I^i} |m_I^i\rangle\langle m_I^i| J_z^i,
\end{equation}
where $m_I^i$ is the magnetic number of the \textit{i}-th nitrogen nucleus ($m_I = 0, \pm 1$ for \ce{^14N} and $m_I = \pm 1/2$ for \ce{^15N}) and $\{|m_I^i\rangle\}$ is the nitrogen dressed basis, i.e., the eigenstates of the hyperfine Hamiltonian in the nitrogen subspace. Figure \ref{F: DEER} shows the simulated coherence of an NV center for a pulsed-DEER, where a $\pi$-pulse is simultaneously applied to the NV and the P1 bath. The frequency of the bath pulse is swept, such that the NV loses coherence when the bath is resonant with the pulse frequency. The dips in Fig. \ref{F: DEER} occur at the resonant frequencies, whose shifts with respect to the electron Larmor frequency match the analytical predictions of Eqs. (\ref{SMEq: 14NHf}) and (\ref{SMEq: 15NHf}).

\subsection{P1 bath Induced Decoherence}

We consider isotopically pure diamond so that there is no presence of \ce{^13C} nuclear spins and the spin bath is composed only of P1 centers. Because we assume a low NV conversion rate, NV centers can be considered independent from each other and therefore the P1 bath remains as the main source of NV dephasing.

\begin{figure}[ht!]
    \centering
    \includegraphics[width=0.5\textwidth]{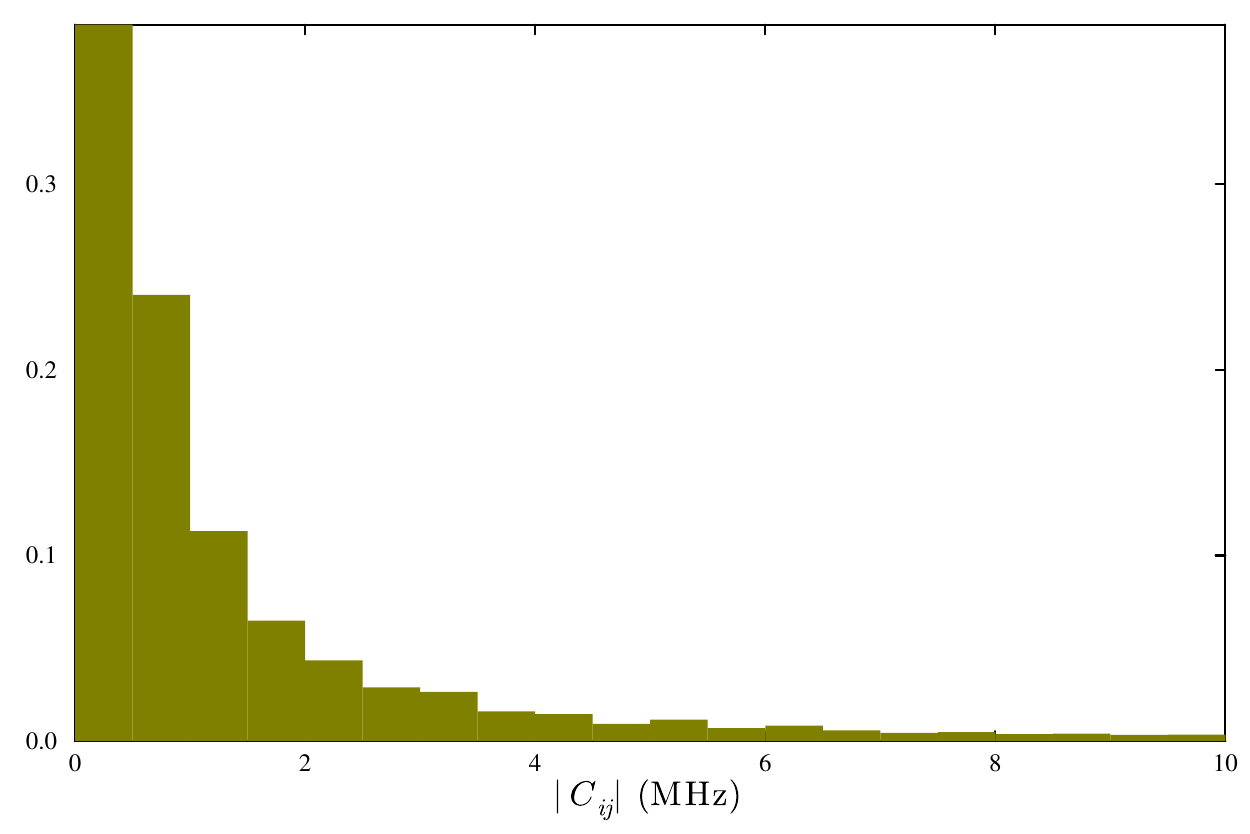}
    \caption{Histogram of the highest coupling strengths between two P1 centers across 100 different spatial configurations of the spin bath at the highest impurity concentration considered in this work, $\rho_{P1} = $ 20 ppm. For lower concentrations, the maximum coupling strengths will show lower values. Because these values are well below the minimum splitting of P1 energies, we can safely assume that any two P1 centers that occupy different energy levels will not experience flip-flopping.}
    \label{F: MaxCoupHist}
\end{figure}

The P1 bath is an electron bath with a hyperfine structure. The energy splitting introduced by the hyperfine interaction results in the suppression of the flip-flopping between P1 centers that occupy different energy levels, leading to ZZ interaction as in \eqref{SMEq: NoFlipFlop}. This occurs because the hyperfine splittings (see Table \ref{T: hfEnergies}) are much larger than the typical coupling strength. To illustrate the latter, as in \cite{SchatzlePCCE}, we show a histogram of the strongest couplings between P1 centers for the highest impurity concentration considered in this work ($\rho_{P1} = $ 20 ppm) in Fig. \ref{F: MaxCoupHist}, demonstrating that we can safely assume that P1 centers with different hyperfine energies will not flip-flop.

\section{Numerical Methods}\label{SM: NumSims}

\subsection{Computing the coherence function}

For simulations in the manuscript, we estimate the Hahn echo $T_2$ time of an ensemble of NV centers which are coupled to the previously outlined P1 bath. The Hahn echo protocol is detailed in Fig.~\ref{F: System} in the main paper. To reproduce the decoherence of NV ensembles, we simulate up to $n_s$ single NVs coupled to different bath configurations and average the resulting coherent functions. To compute these coherence functions, we first prepare the NV center into a superposition state with a $\pi/2$ pulse, such that if the initial state of the NV center is $|0\rangle$ after the pulse, the state is $|X_+\rangle = (|0\rangle + |1\rangle)/\sqrt{2}$, or with density matrix
\begin{equation}
\rho_{NV} = \frac{1}{2}\begin{pmatrix}
1 & 1 \\
1 & 1
\end{pmatrix}.
\end{equation} This state is then allowed to evolve freely for a time $\tau$, where the dynamics are generated with the propagator constructed as
\begin{equation}
U(\tau) = \exp[-i\mathcal{H} \tau].
\end{equation}
Here, $\mathcal{H}$ contains all bath dynamics and NV interactions with this bath in the interaction picture of the free energy, or explicitly 
\begin{equation}
 \mathcal{H} = \sum_{i < j}\mathcal{H}_{ij}^I  + \mathcal{H}_{SB} + \mathcal{H}_D^I,
\end{equation}
 where $\mathcal{H}_{ij}^I$ is given by Eq.~\eqref{SMEq: FlipFlop} or Eq.~\eqref{SMEq: NoFlipFlop} depending on whether the bath spins occupy the same of different energy states respectively; while $\mathcal{H}_{SB}$ ( $\mathcal{H}_D^I$) is described in Eq.~\eqref{SMEq: SB} (Eq.~\eqref{SMEq: EffD}). Subsequently, an instantaneous $\pi$-pulse is applied to the NV, here taken to be $U_\pi = -i\sigma_x$, which inverts the population dynamics, allowing for refocusing of the NV state after a symmetric time $\tau$. The full propagator for this sequence can be constructed piecewise to be
\begin{equation}
U_H(2\tau) = U(\tau) (-i\sigma_x) U(\tau).
\end{equation}
If we define the initial state of the system to be separable, such that the NV center and the bath are not entangled, the density matrix is constructed as $\rho(0) = \rho_{NV} \otimes \rho_{\mathcal{B}}$, where $\rho_{\mathcal{B}} = \bigotimes_{N} \rho_{P1}$ is the initial state of the P1 bath. Then, we define the coherence of the NV after the Hahn echo sequence as the measurement of the $\sigma_x$, that is
\begin{equation}
\mathcal{L}= 2\langle S_x (2\tau)\rangle = \Tr[\hat{U}_H(2\tau) \rho(0) \hat{U}_H(2\tau) \sigma_x].
\end{equation}
Simulating this quantity for a large system exactly and understanding the dynamics of the bath is a formidable problem. Theoretical approaches to study spin bath dynamics include semiclassical models, where the bath is substituted by a random noise, and quantum models. In the former case applied for P1 baths, the coherence of the central spin is largely underestimated, due to the coherent many body evolution of the bath, as well as the hyperfine interaction of P1 centers and their quantum back action on the sensor, which are not considered. In the latter, the high dimensionality of the system renders its analytical treatment inaccessible. Fortunately, numerical methods based on cluster expansions have been developed, allowing the theoretical study of such complex systems. In particular, the Cluster Correlation Expansion (CCE) \cite{CCE-1,CCE-2} has proven to be an extremely useful tool for studying the decoherence of a central spin due to a surrounding bath of weakly coupled spins, such as a nitrogen-vacancy (NV) center in a bath of \ce{^13C}. For the present case, however, the main source for central spin decoherence are the strong intra-bath interactions between P1 centers, which poses additional challenges. These are discussed in the following.

\subsection{Cluster Correlation Expansion}\label{SM: CCE}

In the CCE methodology, the coherence function of the central spin coupled to a large bath of spins, $\mathcal{L}$, is factorized as the product of all the possible irreducible contributions from independent spin clusters $\mathcal{C}$ within the bath:
\begin{equation}\label{SMEq: CCE}
    \mathcal{L} = \prod_{\mathcal{C}} \tilde{\mathcal{L}}_{\mathcal{C}}, \qquad \tilde{\mathcal{L}}_{\mathcal{C}} = \frac{\mathcal{L}_{\mathcal{C}}}{\prod_{\tilde{\mathcal{C}} \in \mathcal{C}} \tilde{\mathcal{L}}_{\tilde{\mathcal{C}}}}.
\end{equation}
This symbolic factorisation represents the product of the contributions of all possible cluster sizes, where each contribution contains the product of every possible combination of spins that make up clusters of that size. If all terms in the expansion are computed and combined, this construction of the coherence is exact; however, the computation of all terms suffers the original large dimension problem. Interestingly, when considering a weakly coupled spin bath where higher order cluster contributions are sequentially included, convergence of the coherence function in Eq. \eqref{SMEq: CCE} is rapid, and usually the expansion can be truncated at an M that is significantly smaller that the total size of the bath. In that case, the coherence function is approximated as
\begin{equation}
    \mathcal{L} \approx \mathcal{L}^{(M)} = \prod_{\mathcal{C}|dim(\mathcal{C}) \leq M} \tilde{\mathcal{L}}_{\mathcal{C}}.
\end{equation}
This is visualised in Fig. \ref{F: CCE}. For example, the coherence function may be well approximated by spin pair contributions, that is, by a second order expansion truncated at $M = 2$ (CCE-2), such that
\begin{equation}
    \mathcal{L} \approx \mathcal{L^{(\mathrm{2})}} = \prod_i \mathcal{L}_i \prod_{ij} \frac{\mathcal{L}_{ij}}{\mathcal{L}_i \mathcal{L}_j},
\end{equation}
where $\mathcal{L}_i$ is the single-spin contribution from the $i$-th spin and $\mathcal{L}_{ij}$ the spin-pair contribution from the cluster including the $i$-th and $j$-th spins.This treatment may be suitable in weakly coupled or sparse baths in which the number of connections (or average cluster size) between bath spins is low, however, for a strongly coupled or dense bath, higher order correlations are required.

\subsection{Mean-field Averaging}\label{SM: MF}

In strongly coupled baths, the ZZ dipolar interaction between bath spins --like that described in Eq.~\eqref{SMEq: NoFlipFlop}-- introduces a significant magnetic field gradient that shifts their resonance energies. This results in the suppression of the flip-flopping of some pairs, as described in Section \ref{SM: Bath Hamiltonian}. Then, if only spin pair effects are considered, every spin pair is allowed to flip-flop, even though in practice many of them would not, given the energy gradients induced by surrounding spins. This leads to an underestimated coherence time of the central spin \cite{WitzelSpaguetti}. On the other hand, the inclusion of higher order correlations in the expansion may lead to unphysical results ($|\mathcal{L}| > 1$) arising from numerical instabilities when two strongly coupled spins are not included in the same cluster correlation term. Furthermore, during our simulations, we noticed that the addition of spin bath driving leads to the apparition of unphysical effects even at second order of approximation. 

In order to avoid these unphysical results, the bath spins that are not explicitly considered in a particular correlation term are included in a mean-field approximation. This approach consists on treating those spins as static during the simulation and averaging the resulting coherence function $\mathcal{L}_C$ for different initial states. Specifically, the spin operators are taken to be $\langle J^i_z \rangle = \pm 1/2$, where the sign depends on their initial state, and $\langle J^i_x \rangle = \langle J^i_y \rangle = 0$. Then, for a given cluster $\mathcal{C}$, the interaction between the cluster and the mean-field is $ \mathcal{H}_{mf}^{\mathcal{C}} = \sum_{j\notin \mathcal{C}}\sum_{i\in\mathcal{C}} H_{mf}^{ij}$, where
\begin{equation}\label{SMEq: mfHam}
    \mathcal{H}_{mf}^{ij} =\text C_{ij} J_z^i \langle J_z^j \rangle.
\end{equation}
This is obtained by substituting in Eq.~\eqref{SMEq: FlipFlop} the spin operators by their respective expected values. Nevertheless, when the spin bath is driven, and such driving has a component perpendicular to the direction $z$, the static condition set above is no longer satisfied, as collective Rabi oscillations of the bath spins are induced in $\langle J_z \rangle $. This poses a problem, as it prevents the straightforward application of the mean-field approach for the analysis of decoupling schemes.

To address this issue, we present an alternative method to include mean-field effects: when the bath spins are driven, the spin projections along the effective driving direction $P$ can be considered static  --instead of along axis $z$--. That is, we can assume that $\langle J_P \rangle = \pm 1/2$ and $\langle J_Q \rangle = \langle J_{Q^\perp} \rangle = 0$. Then, to reproduce the coupling between a cluster spin $i$ and a mean-field spin $j$, we can replace the spin operators of $j$ in the interaction terms (expressed in the tilted basis) with their respective expected values. The mean-field contribution is different depending on whether $i$ and $j$ occupy the same or different energy level, in particular,
\begin{align}
    \text{SAME ENERGY LEVEL: } \quad\;
    \mathcal{H}_{mf}^{ij} = C_{ij} \left(\cos^2\theta - \tfrac{1}{2}\sin^2\theta\right) J_P^i \langle J_P^j \rangle\\
    \text{DIFFERENT ENERGY LEVEL: } \quad\;
    \mathcal{H}_{mf}^{ij} = C_{ij} \cos\theta_i \cos\theta_j J_P^i \langle J_P^j \rangle
\end{align}

where for the former we substitute the expectation values in Eq.~\eqref{SME: IntHam2}, while for the latter in Eq.~\eqref{SMEq: IntHamII}. These expressions are consistent with Eq. \eqref{SMEq: mfHam}, as if no driving is applied, then $\cos\theta = 1$ and $\sin\theta = 0$. Leveraging this formulation, we achieve simulations of dense P1 baths under continuous driving fields without unphysical effects.  Note that, even if the NV center should interact with the mean-field as well, because it is a static interaction, it will be removed by the Hahn echo sequence and therefore we can neglect it.

Regarding the mean-field averaging strategy, two distinct methods exist: external and internal \cite{WitzelSpaguetti,SchatzlePCCE}. In the external averaging, the coherence function is computed for a given initial state of the bath, and then averaged over many different initial states:
\begin{equation}
    \mathcal{L} = \Big\langle \prod_\mathcal{C} \tilde{\mathcal{L}}_\mathcal{C} \Big\rangle.
\end{equation}
In the internal averaging, each irreducible term is averaged separately, and the coherence is subsequently obtained as their product:
\begin{equation}
    \mathcal{L} = \prod_\mathcal{C} \langle \tilde{\mathcal{L}}_\mathcal{C} \rangle, \qquad \langle \tilde{\mathcal{L}}_\mathcal{C}\rangle = \frac{\langle\mathcal{L}_\mathcal{C}\rangle}{\prod_{\tilde{\mathcal{C}} \in \mathcal{C}} \langle \tilde{\mathcal{L}}_\mathcal{\tilde{\mathcal{C}}}\rangle}.
\end{equation}
The latter has proven to produce a better reduction of unphysical behavior in CCE calculations, and so is the one we adopt in this work. 

\subsection{Partition Cluster Correlation Expansion}\label{SM: pCCE}

\begin{figure}

\subfloat[\label{F: pCCE.a}]{%
  \includegraphics[width=0.7\columnwidth]{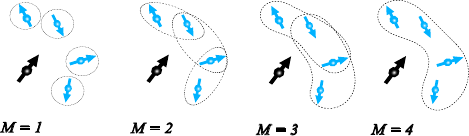}%
}\\
\subfloat[\label{F: pCCE.b}]{%
  \includegraphics[width=0.4\columnwidth]{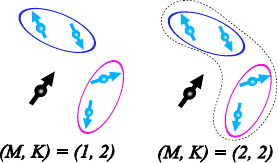}%
}

\caption{A pictorial comparison between (a) CCE-M and (b) pCCE(M, K) for a four spin bath. The central spin is depicted in black while bath spins are shown in light blue. Each cluster is encircled by dashed lines and considered independent from the rest. In every simulation, the central spin is considered together with each cluster. When using the CCE method, the fourth order approximation is needed to include the full system (note that, in low order terms, some clusters may be neglected if their interaction coupling is small enough, for instance, if the spins are far from each other). However, the pCCE method, when partitioning the bath in pairs (the groups circled by blue and magenta lines), already includes the most relevant spin pair correlations at first order and the full system at second order.} \label{F: CCE}
\end{figure}

Although CCE methods have proven to be successful in simulating NV center decoherence \cite{ParkNature,Park2025NVDecoherence,Ghassemizadeh2024,Bauch2020}, they suffer from inefficiency when the average cluster size becomes large, i.e. in strongly coupled baths. To address this, Schäztle \textit{et al.} have recently developed a modification of the conventional CCE method, partition CCE (pCCE) \cite{SchatzlePCCE}, where the bath is first partitioned into groups of strongly interacting spins via a k-Means algorithm and then the CCE-$M$ method is applied onto partitions of size $K$, denoted pCCE($M$,$K$). This means that first order correlations (CCE-1) will now contain strongly coupled $K$ spins instead of individual spins. This avoids the separation of strongly coupled spins into different correlation terms, thus reducing the probability that numerical instabilities appear and consequently the required number of mean-field averages. In fact, further reduction could be obtained by increasing the partition size $K$, but for efficiency, a balance between $K$ and the number of mean field averages should be met. Furthermore, the pCCE method provides an improved convergence of the approximated coherence to the exact result. This is because, at the same order of approximation, the pCCE method also includes the most relevant contributions of higher order. For instance, first order CCE includes single spin correlation while first order pCCE, with a partition size $K=2$, also includes the most relevant spin pair correlations (see Fig. \ref{F: CCE}). For a more thorough insight on the differences between CCE and pCCE, we refer the reader to the Supplementary Material of Ref. \cite{SchatzlePCCE}. As in this study, we find that for NV centers in a P1 bath, an efficient choice of partition size is pCCE(2,4).

\subsection{Details on the numerical simulations}\label{SM: Sims}

\begin{figure}[ht!]
    \centering
    \includegraphics[width=0.5\linewidth]{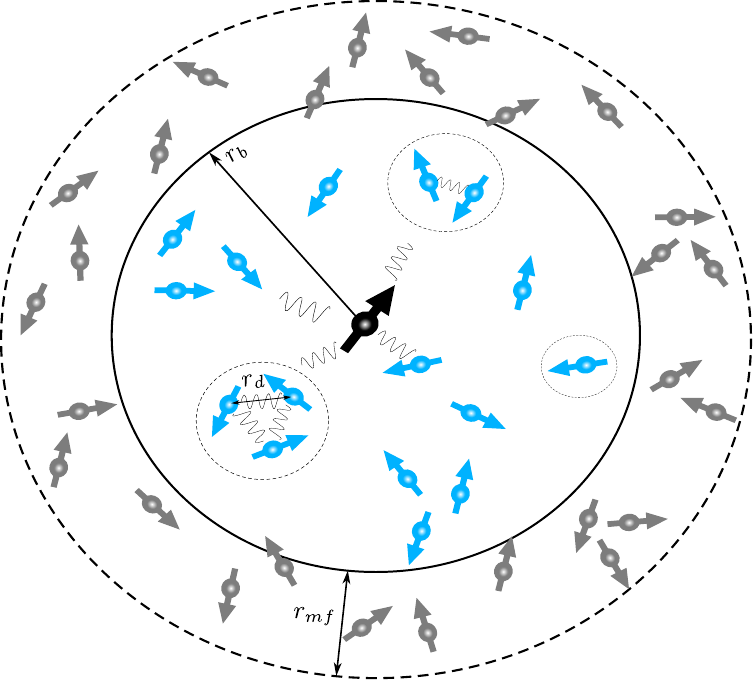}
    \caption{Pictorial representation of the simulated system. Bath spins (in light blue) are those inside a sphere of radius $r_b$. These form clusters consisting of nearby spins and only interactions between those closer than a distance $r_d$ are considered. Mean-field only spins (in gray) are those enclosed in the sell between $r_b$ and $r_{mf}$.}
    \label{F: SMSystem}
\end{figure}

To simulate the NV coherence, we first generate a large diamond lattice with a single NV center at the origin and populate each lattice site with P1 centers with a probability determined by their concentration. Only those impurities within the sphere defined by radius $r_b$ are considered in the bath, while those that lie in the shell defined by radius $r_b$ and $r_{mf}$ ($r_b<r_{mf}$ ) are included in the calculation only at mean-field level --refer to Fig. \ref{F: SMSystem} for an illustrative scheme--. We set $r_b$ ($r_{mf}$) such that there are 180 ($\approx$ 7000) bath (mean-field only) spins. Next, the bath is partitioned into subsystems of four strongly interacting spins ($K=4$) and apply a second order CCE over the partitions, that is pCCE(2,4). We neglect the interactions between P1 centers that are far apart; that is, we define a cutoff distance $r_d$ such that any pair of P1 centers separated by more than $r_d$ is regarded as non-interacting. As in \cite{SchatzlePCCE}, we set $r_d = 65 \mathrm{\ nm} / \rho_{P1} \mathrm{(ppm)}$. To account for the mean field contribution, for a given spatial configuration of the bath spins, we carry out internal averaging over 100 initial states of the mean-field spins. Finally, the ensemble coherence is obtained by averaging the coherence function corresponding to $n_s=50$ distinct spatial arrangements of the bath.

The largest clusters considered in the pCCE(2, 4) approximation contain 9 spins, 8 from the bath plus the NV center, which is included in every cluster. Therefore, the matrices describing the Hamiltonian are significantly large, but fortunately they are also vastly sparse, which allows the use of sparse matrices that reduce the computational time of the simulations. On top of that, for an efficient computation, we perform our simulations in the Hyperion Supercomputer of the Donostia International Physics Center.  

\section{Fitting the Data}\label{SM: Fitting}

\begin{figure}[ht!]
    \centering
    \includegraphics[width=0.95\linewidth]{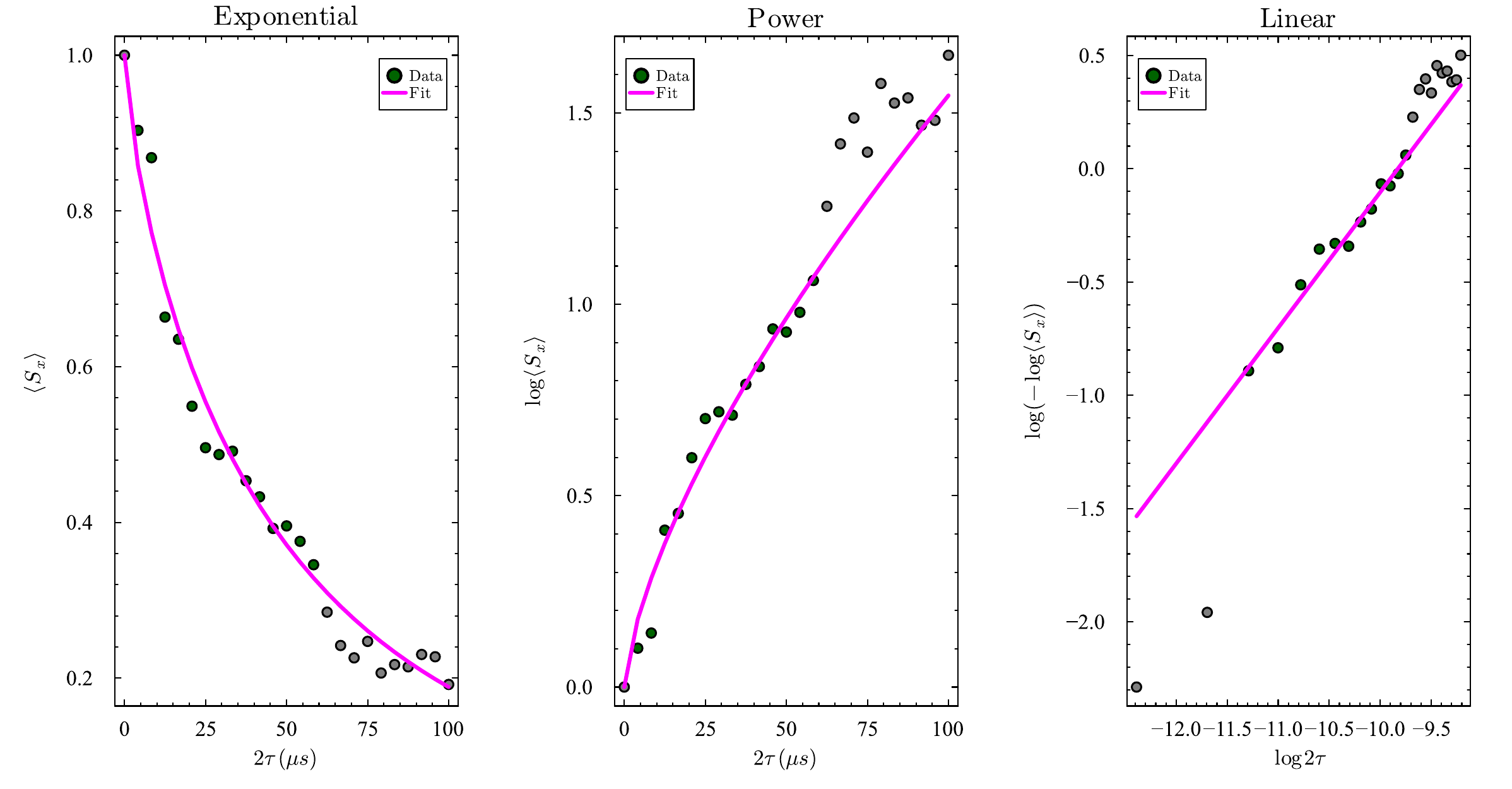}
    \caption{Exponential, power and linear fitting of the NV coherence signal for a P1 concentration $\rho_{P1} = 20$ ppm under our Hybrid-LG bath decoupling protocol ($\Omega = 7$MHz). Data points (not) considered for the fitting are shown in green (gray).}
    \label{F: MixFittings}
\end{figure}

For the extraction of the coherence time of the NV center, we fit the obtained coherence function to a stretched-exponential  $\mathcal{L} = e^{-(2\tau / T_2) ^p}$. We acknowledge that the specific fitting applied to the numerical data, might have an influence on both: the coherence time of the NV center and the stretched exponential parameter \cite{Ghassemizadeh2024}. There are two reasons for this. On the one hand, the decay of the NV coherence is well-reproduced by an stretched exponential function only during the early stages. Consequently, only the initial fraction of the computed signal should be employed for the fitting, and the precise choice of the fraction may slightly influence the extracted parameters. On the other hand, there are several fitting methods: the data can be fitted directly to the stretched exponential, which is the most straightforward approach; alternatively, one could fit the logarithm of the data to a power law (power fit)
\begin{equation*}
    -\log\mathcal{L} \approx \left(\frac{2\tau}{T_2}\right)^p,
\end{equation*}
or taking again the logarithm, via a linear fit
\begin{equation*}
    \log\left(-\log\mathcal{L}\right) \approx p\log2\tau - p\log T_2.
\end{equation*}
In our case, we consider 60\% of the computed signal for extracting both the coherence time and the stretched parameter of the decay. Also, we neglect data points that strongly deviate from the expected behavior. We note that these are not common and can be further suppressed by increasing the number of internal averages performed in the numerical calculations. Figure \ref{F: MixFittings} shows the exponential, power and linear fitting for the coherence signal corresponding to an ensemble of NV centers in a P1 bath of concentration of 20 ppm and under our Hybrid-LG bath decoupling protocol ($\Omega = 7$MHz). The fitted values are displayed in Table \ref{T: MixFits}. We observe that, while the fitted coherence time is similar in all three cases, the stretched parameter is not, what exemplifies the disparities that may arise just from the chosen fitting method. 

\begin{table}[ht!]
    \centering
    \caption{Coherence time and stretched parameter for the NV coherence decay in a P1 bath of concentration of 20 ppm and under our Hybrid-LG bath decoupling protocol ($\Omega = 7$MHz).}
    \begin{tabular}{l|ll}
        \hline
        \hline
        Type of fitting & $T_2 \ (\mathrm{\mu s})$ & $p$ \\
        \hline
        Exponential & 50.59 & 0.75 \\
        Power & 52.75 & 0.68 \\
        Linear & 53.96 & 0.60 \\
        \hline
        \hline
    \end{tabular}
    \label{T: MixFits}
\end{table}

\begin{figure}[ht!]
    \centering
    \includegraphics[width=0.95\linewidth]{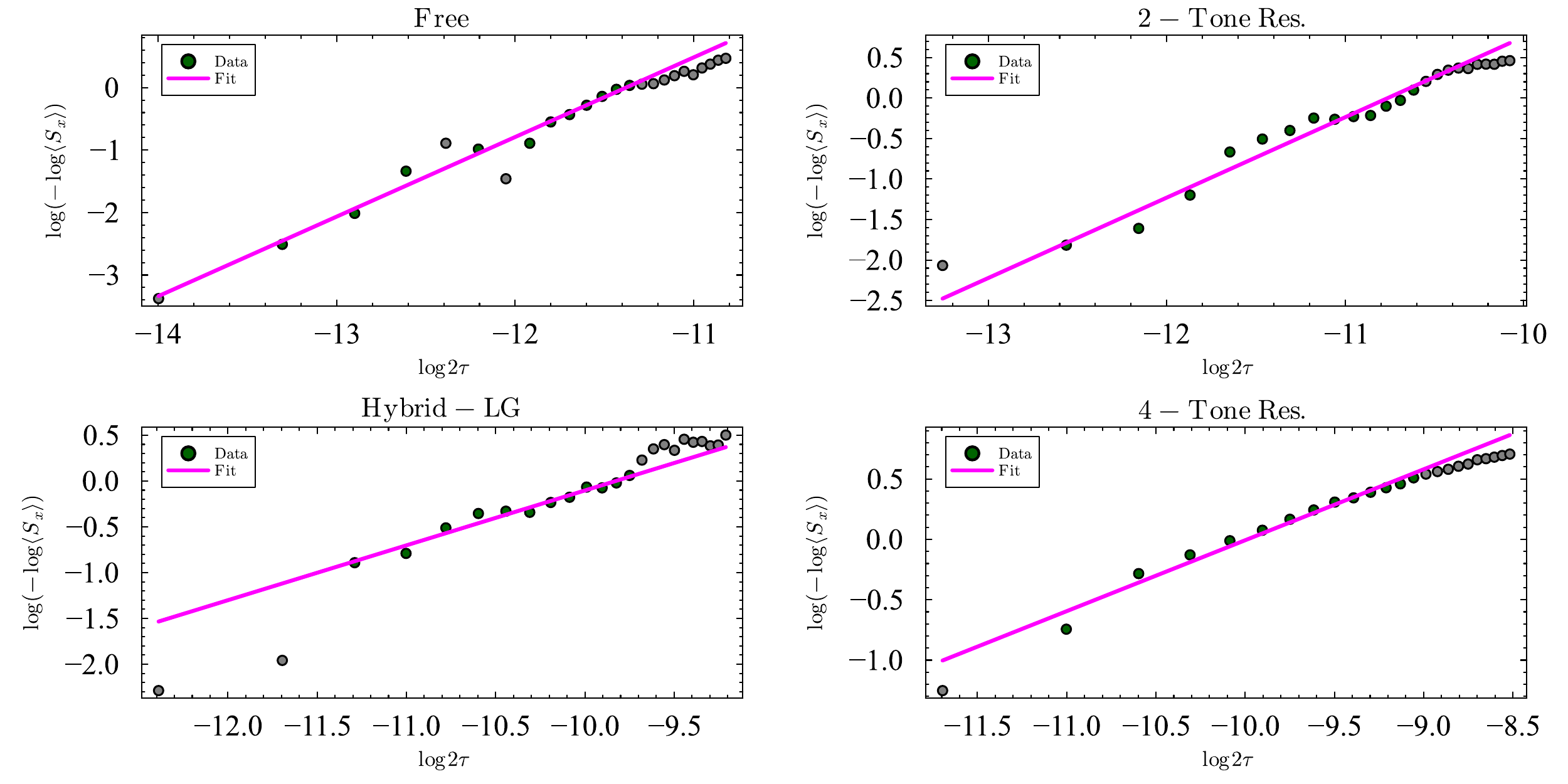}
    \caption{Linear fitting of the NV coherence signal for a P1 concentration of $\rho_{P1} = 20$ ppm under the four bath decoupling protocols considered in the main text (no bath driving, two-tone resonant driving, two-tone Hybrid-LG and four-tone resonant driving), with a Rabi frequency of $\Omega = 7$MHz per tone. Data points (not) considered for the fitting are shown in green (gray).}
    \label{F: LinearFitting}
\end{figure}

For the results in the main text, we have chosen to keep the values obtained via linear fitting, as in common in literature. In Fig. \ref{F: LinearFitting} we show the fitted signal for the four bath decoupling techniques considered in the main text for a P1 concentration of $\rho_{P1} = 20$ ppm and a Rabi frequency of $\Omega = 7$ MHz (per driving). The fitted values are shown in Table \ref{T: LinearFits}.

\begin{table}[ht!]
    \centering
    \caption{Coherence time and stretched parameter for the NV coherence decay in a P1 bath of concentration of 20 ppm and under the four bath decoupling protocols considered in the main text (no bath driving, two-tone resonant driving, two-tone Hybrid-LG and four-tone resonant driving), with a Rabi frequency $\Omega = 7$MHz per tone.}
    \begin{tabular}{l|ll}
        \hline
        \hline
        Type of bath driving & $T_2 \ (\mu s)$ & $p$ \\
        \hline
        None (Free) & 11.43 & 1.27 \\
        Two-tone Resonant & 21.20 & 0.99 \\
        Hybrid-LG & 53.96 & 0.60 \\
        Four-tone Resonant & 45.97 & 0.59 \\
        \hline
        \hline
    \end{tabular}
    \label{T: LinearFits}
\end{table}

\newpage

\section{Error In Averaging}

\begin{figure}
\includegraphics[width = 1\linewidth]{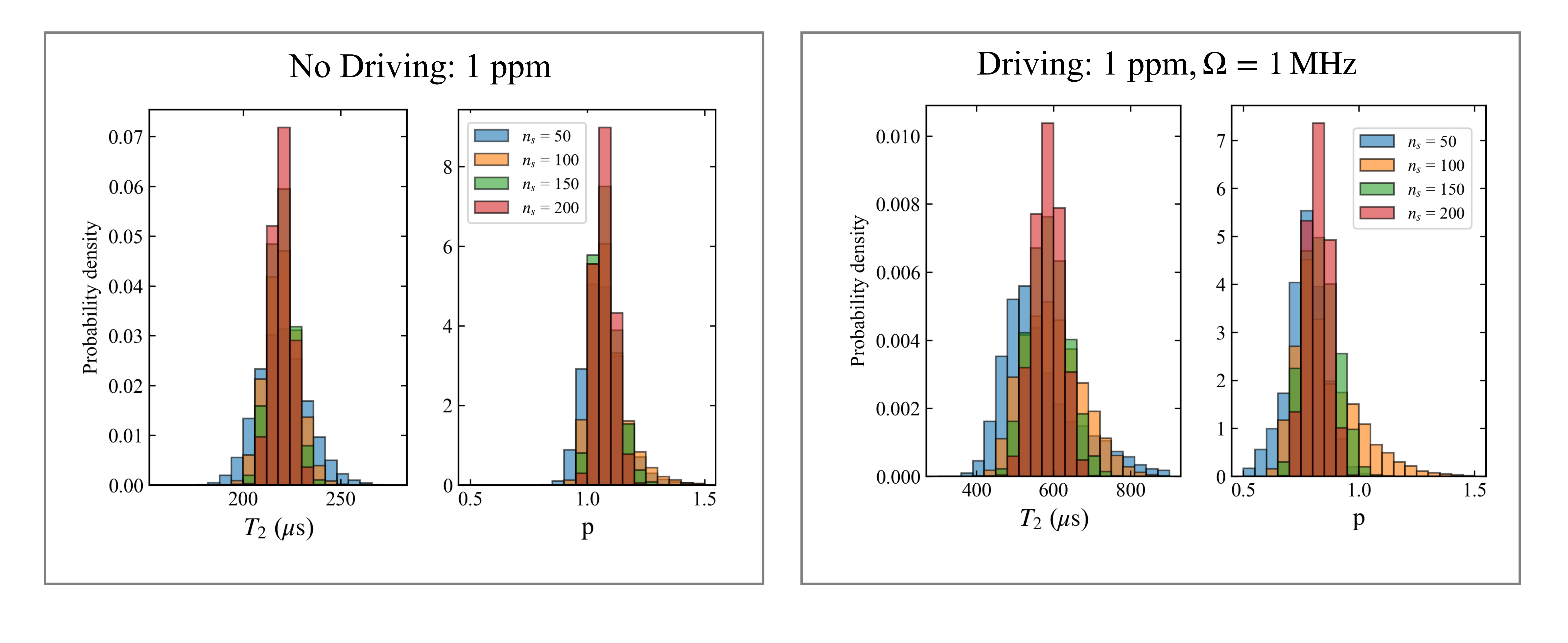}
\caption{Histogram of fitted parameters, $T_2$ and $p$, from a signal constructed with the average of $N$ signals randomly sampled 20000 times from a dataset of 5000 spatial configurations. Different colored histograms represent different size sample average, taking the range $N$ = 50, 100, 150, 200. \textbf{Left panel}: Distribution of signal fitted parameters for the non-driven case of a bath with concentration $\rho_{P1} = 5\,\mathrm{ppm}$. \textbf{Right Panel}: Distribution of signal fitted parameter for the standard driving case, with Rabi frequency $\Omega = 1\,\mathrm{MHz}$ for the sample bath with concentration $\rho_{P1} = 1\,\mathrm{ppm}$.}
\label{F: Error}
\end{figure}

\begin{table}[ht!]
    \centering
    \caption{A table of standard deviations, or errors, in the fitted parameters $T_2$ and $p$ for $\rho_\mathrm{P1} = 1\,\mathrm{ppm}$ extracted from the histograms in Fig. \ref{F: Error}. The errors for both a non-driven and driven bath are presented. The errors are scaled by the mean and presented as a percentage. }
    \begin{tabular}{l|cc|cc}
        \hline
        \hline
         & \multicolumn{2}{c|}{Free} & \multicolumn{2}{c}{Driven} \\
        \hline
        \hline
        $n_s$ & $\sigma_{T_2}\ (\%)$ & $\sigma_p\ (\%)$ & $\sigma_{T_2}\ (\%)$ & $\sigma_p\ (\%)$ \\
        \hline
        50  & 5.8 & 8.9 & 19.8 & 10.9 \\
        100 & 3.7 & 7.9 & 12.6 & 16.2\\
        150 & 2.9 & 4.9 & 8.4 & 8.6 \\
        200 & 2.3 & 3.8 & 6.1 & 5.9 \\
        \hline
    \end{tabular}
    \label{T: Errors}
\end{table}

During our investigation we found that the dominant source of uncertainty in the fitted parameters $T_2$ and $p$ arises from the random sampling of spatial configurations of central spins (NV centers). The ensemble-averaged coherence depends on the specific configurations sampled, and the resulting statistical error decreases approximately as $1/\sqrt{n_s}$ with the number of configurations averaged. In this section we quantify the resulting uncertainty in the extracted parameters for both driven and undriven baths.

For a fixed P1 concentration of $\rho_\mathrm{P1} = 1\,\mathrm{ppm}$, we compute the NV coherence for 5000 independent spatial configurations of the P1 bath. From this set, we draw $n_s$ configurations at random, construct the averaged coherence, and extract $T_2$ and $p$ using fitting procedure in the previous section. Repeating this procedure $2\times 10^4$ times yields distributions for the fitted parameters, which we bin into histograms, as shown in Fig. \ref{F: Error}. We perform this analysis for several values of $n_s$ and for both driven and undriven baths. In all cases, the histogram widths decrease with increasing $N$, reflecting the reduction of statistical uncertainty; the corresponding standard deviations are listed in Table \ref{T: Errors}. Due to the computational expense of including many spatial averages, we settle for $n_s = 50$ for all data in this paper, where the errors are taken from the relevant entry in Table \ref{T: Errors}. These percentages are those used in the errorbars of Fig .3 in the main text.

\section{Low power driving}

\begin{figure}[ht!]
    \centering
    \includegraphics[width=0.5\linewidth]{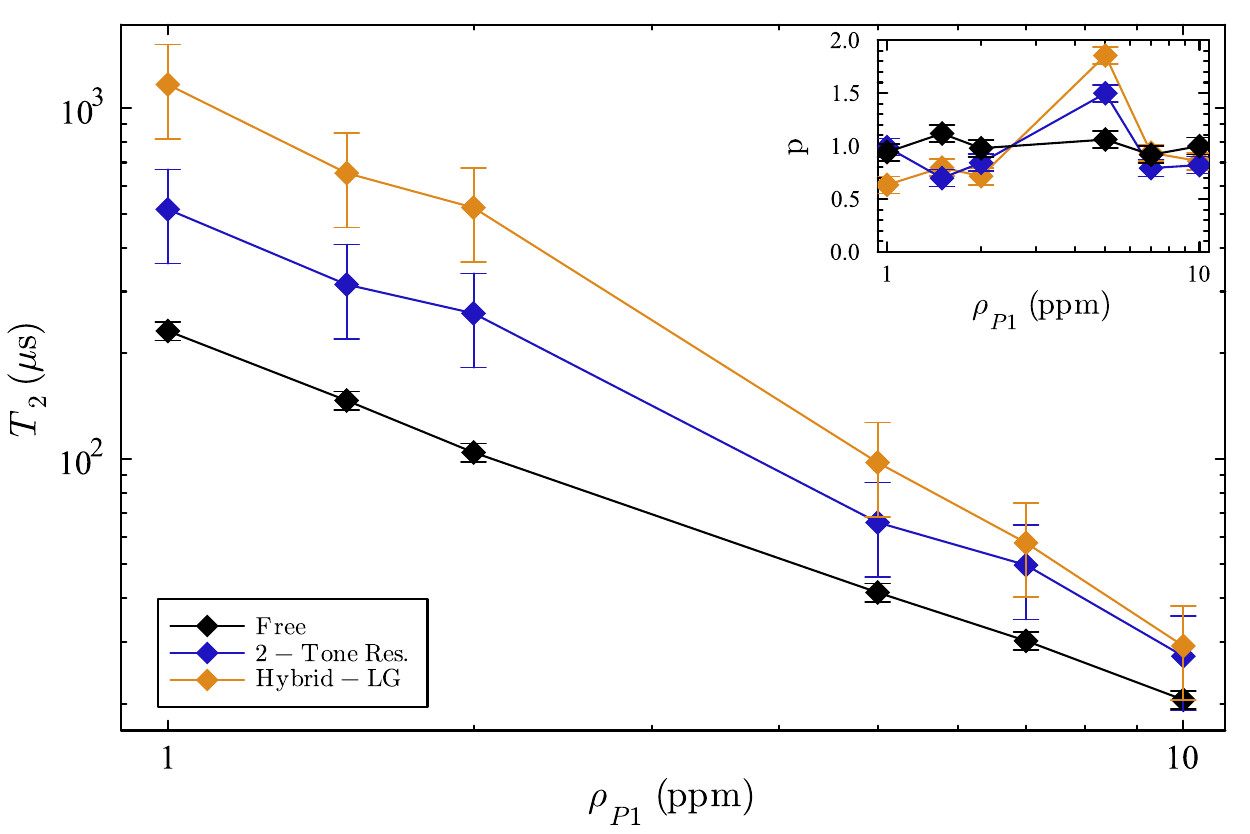}
    \caption{Hahn echo coherence time $T_2$ of an ensemble of NV centers in diamond as a function of the P1 concentration ($\rho_{P1}$) for different spin bath driving protocols with $\Omega=1$ MHz. The same simulation parameters as in Fig.\ref{F: P1Bath} are employed here. For low concentrations, both the two-tone resonant driving (blue) and our Hybrid-LG protocol (orange) yield significant coherence enhancement with respect to no bath driving (black), with Hybrid-LG performing remarkably better.}
    \label{F: log_log_linear_1MHz}
\end{figure}

Throughout this work, we have considered a moderate Rabi frequency, specifically, $\Omega = 7$ MHz per driving. In this section, we compare the two-tone resonant bath driving approach with our Hybrid-LG protocol for a lower Rabi frequency, that is, $\Omega = 1$ MHz. Figure \ref{F: log_log_linear_1MHz} shows the Hahn echo coherence time of an ensemble of NV centers as a function of the P1 concentration. The inset shows the stretched-exponential parameter characterizing the coherence decay for the same concentrations. Our results indicate that, for low driving powers, both protocols fail to meet the conditions necessary for noise suppression at large P1 concentrations, leading to almost no enhancement. However, at lower concentrations, our Hybrid-LG protocol (in orange) yields two-fold improvement with respect to two-tone resonant driving (in blue); thus, even for low Rabi frequencies, significant coherence enhancement can be obtained with our protocol, without additional power consumption.

%% file: lib.bib
@article{ParkNature,
    title = {Decoherence of nitrogen-vacancy spin ensembles in a nitrogen electron-nuclear spin bath in diamond},
    author = {Park, Huijin and Lee, Junghyun and Han, Sangwook and Oh, Sangwon and Seo, Hosung},
    journal = {npj Quantum Information},
    volume = {8},
    issue = {1},
    year = {2022},
    month = {August},
    pages = {95},
    url = {https://doi.org/10.1038/s41534-022-00605-4}
}

@article{MarcksCCE,
  title = {Guiding diamond spin qubit growth with computational methods},
  author = {Marcks, Jonathan C. and Onizhuk, Mykyta and Delegan, Nazar and Wang, Yu-Xin and Fukami, Masaya and Watts, Maya and Clerk, Aashish A. and Heremans, F. Joseph and Galli, Giulia and Awschalom, David D.},
  journal = {Phys. Rev. Mater.},
  volume = {8},
  issue = {2},
  pages = {026204},
  numpages = {11},
  year = {2024},
  month = {Feb},
  publisher = {American Physical Society},
  doi = {10.1103/PhysRevMaterials.8.026204},
  url = {https://link.aps.org/doi/10.1103/PhysRevMaterials.8.026204}
}

@article{SchatzlePCCE,
  title = {Spin coherence in strongly coupled spin baths in quasi-two-dimensional layers},
  author = {Sch\"atzle, Philip and Ghassemizadeh, Reyhaneh and Urban, Daniel F. and Wellens, Thomas and Knittel, Peter and Reiter, Florentin and Jeske, Jan and Hahn, Walter},
  journal = {Phys. Rev. B},
  volume = {110},
  issue = {22},
  pages = {L220302},
  numpages = {7},
  year = {2024},
  month = {Dec},
  publisher = {American Physical Society},
  doi = {10.1103/PhysRevB.110.L220302},
  url = {https://link.aps.org/doi/10.1103/PhysRevB.110.L220302}
}

@misc{Park2025NVDecoherence,
      title={Quantum decoherence of nitrogen-vacancy spin ensembles in a nitrogen spin bath in diamond under dynamical decoupling}, 
      author={Huijin Park and Mykyta Onizhuk and Eunsang Lee and Harim Lim and Junghyun Lee and Sangwon Oh and Giulia Galli and Hosung Seo},
      year={2025},
      eprint={2503.05404},
      archivePrefix={arXiv},
      primaryClass={quant-ph},
      url={https://arxiv.org/abs/2503.05404}, 
}

@article{WitzelSpaguetti,
  title = {Quantum decoherence of the central spin in a sparse system of dipolar coupled spins},
  author = {Witzel, Wayne M. and Carroll, Malcolm S. and Cywi\ifmmode \acute{n}\else \'{n}\fi{}ski, \L{}ukasz and Das Sarma, S.},
  journal = {Phys. Rev. B},
  volume = {86},
  issue = {3},
  pages = {035452},
  numpages = {27},
  year = {2012},
  month = {Jul},
  publisher = {American Physical Society},
  doi = {10.1103/PhysRevB.86.035452},
  url = {https://link.aps.org/doi/10.1103/PhysRevB.86.035452}
}

@article{onizhukCompPersp,
  title = {Colloquium: Decoherence of solid-state spin qubits: A computational perspective},
  author = {Onizhuk, Mykyta and Galli, Giulia},
  journal = {Rev. Mod. Phys.},
  volume = {97},
  issue = {2},
  pages = {021001},
  numpages = {23},
  year = {2025},
  month = {Apr},
  publisher = {American Physical Society},
  doi = {10.1103/RevModPhys.97.021001},
  url = {https://link.aps.org/doi/10.1103/RevModPhys.97.021001}
}

@article{Bauch2020,
  title = {Decoherence of ensembles of nitrogen-vacancy centers in diamond},
  author = {Bauch, Erik and Singh, Swati and Lee, Junghyun and Hart, Connor A. and Schloss, Jennifer M. and Turner, Matthew J. and Barry, John F. and Pham, Linh M. and Bar-Gill, Nir and Yelin, Susanne F. and Walsworth, Ronald L.},
  journal = {Phys. Rev. B},
  volume = {102},
  issue = {13},
  pages = {134210},
  numpages = {9},
  year = {2020},
  month = {Oct},
  publisher = {American Physical Society},
  doi = {10.1103/PhysRevB.102.134210},
  url = {https://link.aps.org/doi/10.1103/PhysRevB.102.134210}
}

@article{knowles-bathDriving,
  author = {Helena S. Knowles and Dhiren M. Kara and Mete Atat\"ure},
  journal = {Nature Materials},
  pages = {21--25},
  title = {Observing bulk diamond spin coherence in high-purity nanodiamonds},
  volume = {13},
  year = {2014},
  url = {https://doi.org/10.1038/nmat3805}
}

@article{Bauch2018,
  title = {Ultralong Dephasing Times in Solid-State Spin Ensembles via Quantum Control},
  author = {Bauch, Erik and Hart, Connor A. and Schloss, Jennifer M. and Turner, Matthew J. and Barry, John F. and Kehayias, Pauli and Singh, Swati and Walsworth, Ronald L.},
  journal = {Phys. Rev. X},
  volume = {8},
  issue = {3},
  pages = {031025},
  numpages = {11},
  year = {2018},
  month = {Jul},
  publisher = {American Physical Society},
  doi = {10.1103/PhysRevX.8.031025},
  url = {https://link.aps.org/doi/10.1103/PhysRevX.8.031025}
}

@article{Bartling2025,
  title = {Control of individual electron-spin pairs in an electron-spin bath},
  author = {Bartling, H. P. and Demetriou, N. and Zutt, N. C. F. and Kwiatkowski, D. and Degen, M. J. and Loenen, S. J. H. and Bradley, C. E. and Markham, M. and Twitchen, D. J. and Taminiau, T. H.},
  journal = {Phys. Rev. Res.},
  volume = {7},
  issue = {1},
  pages = {013333},
  numpages = {6},
  year = {2025},
  month = {Mar},
  publisher = {American Physical Society},
  doi = {10.1103/PhysRevResearch.7.013333},
  url = {https://link.aps.org/doi/10.1103/PhysRevResearch.7.013333}
}

@article{DegenDarkSpins,
  author = {M. J. Degen and S. J. H. Loenen and H. P. Bartling and C. E. Bradley and A. L. Meinsma and M. Markham and D. J. Twitchen and T. H. Taminiau},
  journal = {Nature Communications},
  pages = {3470},
  title = {Entanglement of dark electron-nuclear spin defects in diamond},
  volume = {12},
  year = {2021},
  url = {https://doi.org/10.1038/s41467-021-23454-9}
}

@article{Barry2020sensitivity,
  title = {Sensitivity optimization for NV-diamond magnetometry},
  author = {Barry, John F. and Schloss, Jennifer M. and Bauch, Erik and Turner, Matthew J. and Hart, Connor A. and Pham, Linh M. and Walsworth, Ronald L.},
  journal = {Rev. Mod. Phys.},
  volume = {92},
  issue = {1},
  pages = {015004},
  numpages = {68},
  year = {2020},
  month = {Mar},
  publisher = {American Physical Society},
  doi = {10.1103/RevModPhys.92.015004},
  url = {https://link.aps.org/doi/10.1103/RevModPhys.92.015004}
}

@article{Lee-Goldburg,
  title = {Nuclear-Magnetic-Resonance Line Narrowing by a Rotating rf Field},
  author = {Lee, Moses and Goldburg, Walter I.},
  journal = {Phys. Rev.},
  volume = {140},
  issue = {4A},
  pages = {A1261--A1271},
  numpages = {0},
  year = {1965},
  month = {Nov},
  publisher = {American Physical Society},
  doi = {10.1103/PhysRev.140.A1261},
  url = {https://link.aps.org/doi/10.1103/PhysRev.140.A1261}
}

@article{CCE-1,
  title = {Quantum many-body theory of qubit decoherence in a finite-size spin bath},
  author = {Yang, Wen and Liu, Ren-Bao},
  journal = {Phys. Rev. B},
  volume = {78},
  issue = {8},
  pages = {085315},
  numpages = {13},
  year = {2008},
  month = {Aug},
  publisher = {American Physical Society},
  doi = {10.1103/PhysRevB.78.085315},
  url = {https://link.aps.org/doi/10.1103/PhysRevB.78.085315}
}

@article{CCE-2,
  title = {Quantum many-body theory of qubit decoherence in a finite-size spin bath. II. Ensemble dynamics},
  author = {Yang, Wen and Liu, Ren-Bao},
  journal = {Phys. Rev. B},
  volume = {79},
  issue = {11},
  pages = {115320},
  numpages = {7},
  year = {2009},
  month = {Mar},
  publisher = {American Physical Society},
  doi = {10.1103/PhysRevB.79.115320},
  url = {https://link.aps.org/doi/10.1103/PhysRevB.79.115320}
}

@article{Ghassemizadeh2024,
  title = {Coherence properties of NV-center ensembles in diamond coupled to an electron-spin bath},
  author = {Ghassemizadeh, Reyhaneh and K\"orner, Wolfgang and Urban, Daniel F. and Els\"asser, Christian},
  journal = {Phys. Rev. B},
  volume = {110},
  issue = {20},
  pages = {205148},
  numpages = {9},
  year = {2024},
  month = {Nov},
  publisher = {American Physical Society},
  doi = {10.1103/PhysRevB.110.205148},
  url = {https://link.aps.org/doi/10.1103/PhysRevB.110.205148}
}

@article{Barry2024,
  title = {Sensitive ac and dc magnetometry with nitrogen-vacancy-center ensembles in diamond},
  author = {Barry, John F. and Steinecker, Matthew H. and Alsid, Scott T. and Majumder, Jonah and Pham, Linh M. and O'Keeffe, Michael F. and Braje, Danielle A.},
  journal = {Phys. Rev. Appl.},
  volume = {22},
  issue = {4},
  pages = {044069},
  numpages = {40},
  year = {2024},
  month = {Oct},
  publisher = {American Physical Society},
  doi = {10.1103/PhysRevApplied.22.044069},
  url = {https://link.aps.org/doi/10.1103/PhysRevApplied.22.044069}
}

@misc{AinitzeNVs,
    author = "Biteri-Uribarren, Ainitze and Martin, Ana and Casanova, Jorge",
    title = "{Microscale Sensing with Strongly Interacting NV Ensembles at High Fields}",
    eprint = "2410.21182",
    archivePrefix = "arXiv",
    primaryClass = "quant-ph",
    month = "10",
    year = "2024"
}

@article{DegenCapellaroReview,
  title = {Quantum sensing},
  author = {Degen, C. L. and Reinhard, F. and Cappellaro, P.},
  journal = {Rev. Mod. Phys.},
  volume = {89},
  issue = {3},
  pages = {035002},
  numpages = {39},
  year = {2017},
  month = {Jul},
  publisher = {American Physical Society},
  doi = {10.1103/RevModPhys.89.035002},
  url = {https://link.aps.org/doi/10.1103/RevModPhys.89.035002}
}

@article{Zwerver_2023,
  title = {Shuttling an Electron Spin through a Silicon Quantum Dot Array},
  author = {Zwerver, A.M.J. and Amitonov, S.V. and de Snoo, S.L. and Madzik, M.T. and Rimbach-Russ, M. and Sammak, A. and Scappucci, G. and Vandersypen, L.M.K.},
  journal = {PRX Quantum},
  volume = {4},
  issue = {3},
  pages = {030303},
  numpages = {11},
  year = {2023},
  month = {Jul},
  publisher = {American Physical Society},
  doi = {10.1103/PRXQuantum.4.030303},
  url = {https://link.aps.org/doi/10.1103/PRXQuantum.4.030303}
}

@article{CastellettoColorCenters,
doi = {10.1088/2515-7647/ab77a2},
url = {https://dx.doi.org/10.1088/2515-7647/ab77a2},
year = {2020},
month = {mar},
publisher = {IOP Publishing},
volume = {2},
number = {2},
pages = {022001},
author = {Castelletto, Stefania and Boretti, Alberto},
title = {Silicon carbide color centers for quantum applications},
journal = {Journal of Physics: Photonics},
}

@article{BelthangadyNVP1,
  title = {Dressed-{S}tate {R}esonant {C}oupling between {B}right and {D}ark {S}pins in {D}iamond},
  author = {Belthangady, C. and Bar-Gill, N. and Pham, L. M. and Arai, K. and Le Sage, D. and Cappellaro, P. and Walsworth, R. L.},
  journal = {Phys. Rev. Lett.},
  volume = {110},
  issue = {15},
  pages = {157601},
  numpages = {5},
  year = {2013},
  month = {Apr},
  publisher = {American Physical Society},
  doi = {10.1103/PhysRevLett.110.157601},
  url = {https://link.aps.org/doi/10.1103/PhysRevLett.110.157601}
}

@article{ZhaoNV,
  title = {Decoherence and dynamical decoupling control of nitrogen vacancy center electron spins in nuclear spin baths},
  author = {Zhao, Nan and Ho, Sai-Wah and Liu, Ren-Bao},
  journal = {Phys. Rev. B},
  volume = {85},
  issue = {11},
  pages = {115303},
  numpages = {18},
  year = {2012},
  month = {Mar},
  publisher = {American Physical Society},
  doi = {10.1103/PhysRevB.85.115303},
  url = {https://link.aps.org/doi/10.1103/PhysRevB.85.115303}
}

@article{Herbschleb_2019,
  title = {Ultra-long coherence times amongst room-temperature solid-state spins},
  author = {Herbschleb, E. D. and Kato, H. and Maruyama, Y. and Danjo, T. and Makino, T. and Yamasaki, S. and Ohki, I. and Hayashi, K. and Morishita, H. and Fujiwara, M. and Mizuochi, N.},
  journal = {Nature Communications},
  volume = {10},
  issue = {1},
  pages = {3766},
  year = {2019},
  doi = {10.1038/s41467-019-11776-8},
  url = {https://doi.org/10.1038/s41467-019-11776-8}
}

@article{Abobeih_2022,
author = {Abobeih, M. H. and Wang, Y. and Randall, J. and Loenen, S. J. H. and Bradley, C. E. and Markham, M. and Twitchen, D. J. and Terhal, B. M. and Taminiau, T. H.},
title = {Fault-tolerant operation of a logical qubit in a diamond quantum processor},
journal = {Nature},
volume = {606},
number = {7916},
pages = {884--889},
year = {2022},
doi = {10.1038/s41586-022-04819-6},
URL = {https://doi.org/10.1038/s41586-022-04819-6}
}

@article{Huang_2024,
author = {Huang, Jonathan Y. and Su, Rocky Y. and Lim, Wee Han and Feng, MengKe and van Straaten, Barnaby and Severin, Brandon and Gilbert, Will and Dumoulin Stuyck, Nard and Tanttu, Tuomo and Serrano, Santiago and Cifuentes, Jesus D. and Hansen, Ingvild and Seedhouse, Amanda E. and Vahapoglu, Ensar and Leon, Ross C. C. and Abrosimov, Nikolay V. and Pohl, Hans-Joachim and Thewalt, Michael L. W. and Hudson, Fay E. and Escott, Christopher C. and Ares, Natalia and Bartlett, Stephen D. and Morello, Andrea and Saraiva, Andre and Laucht, Arne and Dzurak, Andrew S. and Yang, Chih Hwan},
title = {High-fidelity spin qubit operation and algorithmic initialization above 1 K},
journal = {Nature},
volume = {627},
number = {8005},
pages = {772--777},
year = {2024},
doi = {10.1038/s41586-024-07160-2},
URL = {https://doi.org/10.1038/s41586-024-07160-2}
}

@article{Simmons_2024,
  title = {Scalable Fault-Tolerant Quantum Technologies with Silicon Color Centers},
  author = {Simmons, Stephanie},
  journal = {PRX Quantum},
  volume = {5},
  issue = {1},
  pages = {010102},
  numpages = {18},
  year = {2024},
  month = {Mar},
  publisher = {American Physical Society},
  doi = {10.1103/PRXQuantum.5.010102},
  url = {https://link.aps.org/doi/10.1103/PRXQuantum.5.010102}
}

@article{Du_2024,
  title = {Single-molecule scale magnetic resonance spectroscopy using quantum diamond sensors},
  author = {Du, Jiangfeng and Shi, Fazhan and Kong, Xi and Jelezko, Fedor and Wrachtrup, J\"org},
  journal = {Rev. Mod. Phys.},
  volume = {96},
  issue = {2},
  pages = {025001},
  numpages = {62},
  year = {2024},
  month = {May},
  publisher = {American Physical Society},
  doi = {10.1103/RevModPhys.96.025001},
  url = {https://link.aps.org/doi/10.1103/RevModPhys.96.025001}
}

@article{Hong_2013,
author = {Hong, Sungkun and Grinolds, Michael S. and Pham, Linh M. and Le Sage, David and Luan, Lan and Walsworth, Ronald L. and Yacoby, Amir},
title = {Nanoscale magnetometry with NV centers in diamond},
journal = {MRS Bulletin},
volume = {38},
number = {2},
pages = {155--161},
year = {2013},
doi = {10.1557/mrs.2013.23},
URL = {https://doi.org/10.1557/mrs.2013.23}
}

@article{Hermans_2022,
author = {Hermans, S. L. N. and Pompili, M. and Beukers, H. K. C. and Baier, S. and Borregaard, J. and Hanson, R.},
title = {Qubit teleportation between non-neighbouring nodes in a quantum network},
journal = {Nature},
volume = {605},
number = {7911},
pages = {663--668},
year = {2022},
doi = {10.1038/s41586-022-04697-y},
URL = {https://doi.org/10.1038/s41586-022-04697-y}
}

@article{Joos_2022,
author = {Joos, Maxime and Bluvstein, Dolev and Lyu, Yuanqi and Weld, David and Bleszynski Jayich, Ania},
title = {Protecting qubit coherence by spectrally engineered driving of the spin environment},
journal = {npj Quantum Information},
volume = {8},
number = {1},
pages = {47},
year = {2022},
doi = {10.1038/s41534-022-00560-0},
URL = {https://doi.org/10.1038/s41534-022-00560-0}
}

@article{Chrostoski_2021,
title = {Magnetic field noise analyses generated by the interactions between a nitrogen vacancy center diamond and surface and bulk impurities},
journal = {Physica B: Condensed Matter},
volume = {605},
pages = {412767},
year = {2021},
issn = {0921-4526},
doi = {https://doi.org/10.1016/j.physb.2020.412767},
url = {https://www.sciencedirect.com/science/article/pii/S0921452620307419},
author = {Philip Chrostoski and Bruce Barrios and D.H. Santamore}
}

@article{Zheng_2022,
author = {Zheng, Wentian and Bian, Ke and Chen, Xiakun and Shen, Yang and Zhang, Shichen and Stöhr, Rainer and Denisenko, Andrej and Wrachtrup, Jörg and Yang, Sen and Jiang, Ying},
title = {Coherence enhancement of solid-state qubits by local manipulation of the electron spin bath},
journal = {Nature Physics},
volume = {18},
number = {11},
pages = {1317--1323},
year = {2022},
doi = {10.1038/s41567-022-01719-4},
URL = {https://doi.org/10.1038/s41567-022-01719-4}
}

@article{Hao_2025,
author = {Hao, Yucheng and Yang, Zhiping and Li, Zeyu and Kong, Xi and Tang, Wenna and Xie, Tianyu and Xu, Shaoyi and Ye, Xiangyu and Yu, Pei and Wang, Pengfei and Wang, Ya and Qiao, Zhenhua and Gao, Libo and Jiang, Jian-Hua and Shi, Fazhan and Du, Jiangfeng},
title = {Coherence enhancement via a diamond-graphene hybrid for nanoscale quantum sensing},
journal = {National Science Review},
volume = {12},
number = {5},
year = {2025},
doi = {10.1093/nsr/nwaf076},
URL = {https://doi.org/10.1093/nsr/nwaf076}
}

@article{KiLo_2025,
author = {Lo, Wing Ki and Zhang, Yaowen and Chow, Ho Yin and Wu, Jiahao and Leung, Man Yin and Ho, Kin On and Du, Xuliang and Chen, Yifan and Shen, Yang and Pan, Ding and Yang, Sen},
title = {Enhancement of quantum coherence in solid-state qubits via interface engineering},
journal = {Nature Communications},
volume = {16},
number = {1},
pages = {5984},
year = {2025},
doi = {10.1038/s41467-025-61026-3},
URL = {https://doi.org/10.1038/s41467-025-61026-3}
}

@article{FSLG,
title = {Frequency-switched pulse sequences: Homonuclear decoupling and dilute spin NMR in solids},
journal = {Chemical Physics Letters},
volume = {155},
number = {4},
pages = {341-346},
year = {1989},
issn = {0009-2614},
doi = {https://doi.org/10.1016/0009-2614(89)87166-0},
url = {https://www.sciencedirect.com/science/article/pii/0009261489871660},
author = {A. Bielecki and A.C. Kolbert and M.H. Levitt},
}

@article{Halse_LG4,
title = {Improved Phase-Modulated Homonuclear Dipolar Decoupling for Solid-State NMR Spectroscopy from Symmetry Considerations},
journal = {The Journal of Physical Chemistry A},
volume = {117},
number = {25},
pages = {5280-5290},
year = {2013},
doi = {10.1021/jp4038733},
url = {https://doi.org/10.1021/jp4038733},
author = {Halse, Meghan E. and Emsley, Lyndon},
}

@article{Randall_Simulator,
author = {J. Randall  and C. E. Bradley  and F. V. van der Gronden  and A. Galicia  and M. H. Abobeih  and M. Markham  and D. J. Twitchen  and F. Machado  and N. Y. Yao  and T. H. Taminiau },
title = {Many-body–localized discrete time crystal with a programmable spin-based quantum simulator},
journal = {Science},
volume = {374},
number = {6574},
pages = {1474-1478},
year = {2021},
doi = {10.1126/science.abk0603},
URL = {https://www.science.org/doi/abs/10.1126/science.abk0603},
eprint = {https://www.science.org/doi/pdf/10.1126/science.abk0603}}

@article{Blind_QC,
author = {Y.-C. Wei  and P.-J. Stas  and A. Suleymanzade  and G. Baranes  and F. Machado  and Y. Q. Huan  and C. M. Knaut  and S. W. Ding  and M. Merz  and E. N. Knall  and U. Yazlar  and M. Sirotin  and I. W. Wang  and B. Machielse  and S. F. Yelin  and J. Borregaard  and H. Park  and M. Lončar  and M. D. Lukin },
title = {Universal distributed blind quantum computing with solid-state qubits},
journal = {Science},
volume = {388},
number = {6746},
pages = {509-513},
year = {2025},
doi = {10.1126/science.adu6894},
URL = {https://www.science.org/doi/abs/10.1126/science.adu6894},
eprint = {https://www.science.org/doi/pdf/10.1126/science.adu6894}}

@article{Wolfowicz_2021,
author = {Wolfowicz, Gary and Heremans, F. Joseph and Anderson, Christopher P. and Kanai, Shun and Seo, Hosung and Gali, Adam and Galli, Giulia and Awschalom, David D.},
title = {Quantum guidelines for solid-state spin defects},
journal = {Nature Reviews Materials},
volume = {6},
number = {10},
pages = {906--925},
year = {2021},
doi = {10.1038/s41578-021-00306-y},
URL = {https://doi.org/10.1038/s41578-021-00306-y}}

@article{Liu_2022,
doi = {10.1088/2633-4356/ac7e9f},
url = {https://doi.org/10.1088/2633-4356/ac7e9f},
year = {2022},
month = {jul},
publisher = {IOP Publishing},
volume = {2},
number = {3},
pages = {032002},
author = {Liu, Wei and Guo, Nai-Jie and Yu, Shang and Meng, Yu and Li, Zhi-Peng and Yang, Yuan-Ze and Wang, Zhao-An and Zeng, Xiao-Dong and Xie, Lin-Ke and Li, Qiang and Wang, Jun-Feng and Xu, Jin-Shi and Wang, Yi-Tao and Tang, Jian-Shun and Li, Chuan-Feng and Guo, Guang-Can},
title = {Spin-active defects in hexagonal boron nitride},
journal = {Materials for Quantum Technology}}

@misc{cAERIS,
      title={High-Field NMR Enhanced Sensitivity via Nuclear Spin Locking with NV Centers}, 
      author={Oliver T. Whaites and Jaime García Oliván and Jorge Casanova},
      year={2025},
      eprint={2504.00887},
      archivePrefix={arXiv},
      primaryClass={quant-ph},
      url={https://arxiv.org/abs/2504.00887}, 
}

@article{Carlos_Q_Sim,
  title = {Co-Design quantum simulation of nanoscale NMR},
  author = {Algaba, Manuel G. and Ponce-Martinez, Mario and Munuera-Javaloy, Carlos and Pina-Canelles, Vicente and Thapa, Manish J. and Taketani, Bruno G. and Leib, Martin and de Vega, In\'es and Casanova, Jorge and Heimonen, Hermanni},
  journal = {Phys. Rev. Res.},
  volume = {4},
  issue = {4},
  pages = {043089},
  numpages = {20},
  year = {2022},
  month = {Nov},
  publisher = {American Physical Society},
  doi = {10.1103/PhysRevResearch.4.043089},
  url = {https://link.aps.org/doi/10.1103/PhysRevResearch.4.043089}
}

@article{Jens_2024,
  title = {Extending radiowave frequency detection range with dressed states of solid-state spin ensembles},
  author = {Hermann, Jens C. and Rizzato, Roberto and Bruckmaier, Fleming and Allert, Robin D. and Blank, Aharon and Bucher, Dominik B.},
  journal = {npj Quantum Information},
  volume = {10},
  issue = {1},
  numpages = {103},
  year = {2024},
  doi = {10.1038/s41534-024-00891-0},
  url = {https://doi.org/10.1038/s41534-024-00891-0}
}

@misc{Grafenstein_2025,
      title={Coherent signal detection in the statistical polarization regime enables high-resolution nanoscale NMR spectroscopy}, 
      author={Nick R. von Grafenstein and Karl D. Briegel and Jorge Casanova and Dominik B. Bucher},
      year={2025},
      eprint={2501.02093},
      archivePrefix={arXiv},
      primaryClass={physics.chem-ph},
      url={https://arxiv.org/abs/2501.02093}, 
}

@misc{Rizzato_2025,
      title={Quantum sensing with spin defects in boron nitride nanotubes}, 
      author={Roberto Rizzato and Andrea Alberdi Hidalgo and Linyan Nie and Elena Blundo and Nick R. von Grafenstein and Jonathan J. Finley and Dominik B. Bucher},
      year={2025},
      eprint={2504.16725},
      archivePrefix={arXiv},
      primaryClass={quant-ph},
      url={https://arxiv.org/abs/2504.16725}, 
}

@article{Carlos_DD,
doi = {10.1209/0295-5075/ac0ed1},
url = {https://doi.org/10.1209/0295-5075/ac0ed1},
year = {2021},
month = {jul},
publisher = {EDP Sciences, IOP Publishing and Società Italiana di Fisica},
volume = {134},
number = {3},
pages = {30001},
author = {Munuera-Javaloy, C. and Puebla, R. and Casanova, J.},
title = {Dynamical decoupling methods in nanoscale NMR},
journal = {Europhysics Letters}}

@article{Han_DD,
  title = {Protecting logical qubits with dynamical decoupling},
  author = {Han, Jia-Xiu and Zhang, Jiang and Xue, Guang-Ming and Yu, Haifeng and Long, Guilu},
  journal = {Phys. Rev. Appl.},
  volume = {24},
  issue = {2},
  pages = {024003},
  numpages = {12},
  year = {2025},
  month = {Aug},
  publisher = {American Physical Society},
  doi = {10.1103/lm8x-r48q},
  url = {https://link.aps.org/doi/10.1103/lm8x-r48q}
}

@article{Rizzato_DD,
  title = {Extending the coherence of spin defects in hBN enables advanced qubit control and quantum sensing},
  author = {Rizzato, Roberto and Schalk, Martin and Mohr, Stephan and Hermann, Jens C. and Leibold, Joachim P. and Bruckmaier, Fleming and Salvitti, Giovanna and Qian, Chenjiang and Ji, Peirui and Astakhov, Georgy V. and Kentsch, Ulrich and Helm, Manfred and Stier, Andreas V. and Finley, Jonathan J. and Bucher, Dominik B.},
  journal = {Nature Communications},
  volume = {14},
  issue = {1},
  pages = {5089},
  year = {2025},
  doi = {10.1038/s41467-023-40473-w},
  url = {https://doi.org/10.1038/s41467-023-40473-w}}

@article{Ezzell_DD,
  title = {Dynamical decoupling for superconducting qubits: A performance survey},
  author = {Ezzell, Nic and Pokharel, Bibek and Tewala, Lina and Quiroz, Gregory and Lidar, Daniel A.},
  journal = {Phys. Rev. Appl.},
  volume = {20},
  issue = {6},
  pages = {064027},
  numpages = {42},
  year = {2023},
  month = {Dec},
  publisher = {American Physical Society},
  doi = {10.1103/PhysRevApplied.20.064027},
  url = {https://link.aps.org/doi/10.1103/PhysRevApplied.20.064027}
}

@article{Arunkumar_RR,
  title = {Quantum Logic Enhanced Sensing in Solid-State Spin Ensembles},
  author = {Arunkumar, Nithya and Olsson, Kevin S. and Oon, Jner Tzern and Hart, Connor A. and Bucher, Dominik B. and Glenn, David R. and Lukin, Mikhail D. and Park, Hongkun and Ham, Donhee and Walsworth, Ronald L.},
  journal = {Phys. Rev. Lett.},
  volume = {131},
  issue = {10},
  pages = {100801},
  numpages = {7},
  year = {2023},
  month = {Sep},
  publisher = {American Physical Society},
  doi = {10.1103/PhysRevLett.131.100801},
  url = {https://link.aps.org/doi/10.1103/PhysRevLett.131.100801}
}
